\documentclass{aa} 

\def\dosingle#1::::{#1}  \def\dodouble#1::::{ } 

\dodouble \documentclass[referee]{aa} ::::

\newcommand\dobold[1]{{#1}}

\usepackage{natb_2e}
\usepackage{epsfcopy}

\def\nice#1::::{#1}    \def\subm#1::::{}   

\newcommand\zzz[2]{#2}  

\def\SS{Sect.~}

\def\apj{ApJ}                 
\def\aj{AJ}                       
\def\aap{A\&A}            
\def\aaps{A\&AS}            
\def\mnras{MNRAS}

\renewcommand\citep[1]{(\citealt{#1})}
\newcommand\citepf[1]{(\citealt*{#1})}    

\def\.{{\cdot}} 
\def\gtapprox{\,\lower.6ex\hbox{$\buildrel >\over \sim$} \, }
\def\ltapprox{\,\lower.6ex\hbox{$\buildrel <\over \sim$} \, }
\def\propapprox{\,\lower.6ex\hbox{$\buildrel \propto\over \sim$} \, }

\def\e{ {\scriptstyle \times} 10^}

\def\arcs{\ifmmode {'' }\else $'' $\fi}     
\def\arcm{\ifmmode {' }\else $' $\fi}       
\def\deg{\ifmmode^\circ\else$^\circ$\fi}    

\def\fr7{7$ \hskip -0.9ex \vrule height0.8ex width0.8ex depth-0.73ex
                                                                \hskip0.1ex$}
\def\frtoday{Le\space\number\day\space\ifcase\month\or
  janvier\or f\'evrier\or mars\or avril\or mai\or juin\or
  juillet\or ao\^ut\or septembre\or octobre\or novembre\or d\'ecembre\fi\space \number\year}

\newcommand\joref[5]{#1, #5, {#2 }{#3, } #4}  
 
\newcommand\epref[3]{#1, #3, #2}

\def\basi{Bull.Ast.Soc.India}   %

\def\hMpc{\mbox{$\,h^{-1}\,$Mpc}}

\def\hGpc{\mbox{$h^{-1}\,$Gpc}}

\def\rmin{r_{\mbox{\rm \small min}}}

\def\llss{L_{\mbox{\rm \small LSS}}}

\def\rmax{r^i_{\mbox{\rm \small max}}}

\def\Omm{\Omega_{\mbox{\rm \small m}}}
\def\wQ{w_{\mbox{\rm \small Q}}}
\def\rsim{r_{\mbox{\rm \small sim}}}
\def\robs{r_{\mbox{\rm \small obs}}}
\def\anddd{{\mbox{\rm \ \ and\ \ }}}

\begin{document}


\title{The Cosmological Constant and Quintessence from 
a Correlation Function Comoving Fine Feature
in the 2dF Quasar Redshift Survey}

\author{B.~F.~Roukema\inst{1,2,3}  
\and
 G.~A.~Mamon\inst{4,5}
\and S. Bajtlik\inst{6}
}

\authorrunning{B. F. Roukema et al.}
\titlerunning{Metric Parameters from a 2QZ-10K Comoving Feature}

   \offprints{B. F. Roukema}

 \institute{Inter-University Centre for Astronomy and Astrophysics, 
    Post Bag 4, Ganeshkhind, Pune, 411 007, India 
{\em (boud.roukema@obspm.fr)}
\and
DARC/LUTH, Observatoire de Paris--Meudon, 
5, place Jules Janssen, F-92195 Meudon Cedex, France
\and
University of Warsaw, Krakowskie Przedmie\'scie 26/28, 
00-927 Warsaw, Poland
\and
Institut d'Astrophysique de Paris
(CNRS UPR 341),
98bis Bd Arago, F-75014 Paris, France 
{\em (gam@iap.fr)}
\and
DAEC (CNRS UMR 8631),
Observatoire de Paris-Meudon, 5 place Jules Janssen, F-92195 Meudon Cedex,
France
\and
Nicolaus Copernicus Astronomical Center, 
ul. Bartycka 18, 00-716 Warsaw, Poland {\em (bajtlik@camk.edu.pl)}
}

\date{\frtoday}


\abstract{
Local maxima at characteristic comoving scales have previously been
claimed to exist in the density perturbation spectrum at the 
wavenumber $k =2\pi/\llss$, 
where $\llss\sim 100$--$200${\hMpc} (comoving), at low
redshift ($z \ltapprox 0.4$) for several classes of tracer objects, at
$z\approx 2$ among quasars, and at $z\approx 3$ among Lyman break
galaxies.
Here, this cosmic standard ruler is sought in the 
``10K'' initial release of the 2dF QSO Redshift Survey (2QZ-10K), 
by estimating the spatial two-point autocorrelation functions $\xi(r)$
of the three-dimensional (comoving, spatial) distribution of
the $N=2378$ quasars in the most completely observed and ``covered''
sky regions of the catalogue, over the 
redshift ranges 
$0.6 < z < 1.1$ (``low-$z$''),
$1.1 < z < 1.6$ (``med-$z$'') and
$1.6 < z < 2.2$ (``hi-$z$''). 
Because of the selection method of the survey and sparsity of the data, 
the analysis was done conservatively to avoid non-cosmological artefacts.
(i) Avoiding {\em a priori} estimates of the length scales of features, 
local maxima in $\xi(r)$ are found in 
all three different redshift ranges.
The requirement that 
a local maximum be present in all three redshift ranges
at a fixed comoving length 
scale implies strong,
purely geometric constraints on the local cosmological parameters,
in which case the length scale of the local maximum common 
to the three redshift ranges is $2\llss= 244\pm17${\hMpc}.
(ii) For a standard cosmological constant FLRW model, the matter density 
and cosmological constant are constrained to 
$\Omm= 0.25\pm0.10, \Omega_\Lambda=0.65{\pm0.25} $ (68\% confidence),
$\Omm= 0.25\pm0.15, \Omega_\Lambda=0.60{\pm0.35} $ (95\% confidence),
respectively, {\em from the 2QZ-10K alone}. 
Independently of the type Ia supernovae data, 
the zero cosmological constant model ($\Omega_\Lambda=0$) is rejected at the 
99.7\% confidence level. 
(iii) For an effective quintessence ($\wQ$) model and zero curvature, 
$ \wQ<-0.5 $ (68\% confidence),
$ \wQ<-0.35 $ (95\% confidence) are found,
again {\em from the 2QZ-10K alone}.
\dobold{
In a different analysis of a larger (but
less complete) subset of the same 2QZ-10K catalogue, \citet{Hoyle01}
found a local maximum in the power spectrum 
to exist for widely differing choices of 
$\Omm$ and $\Omega_\Lambda$, which is difficult to understand for 
a genuine large scale feature at fixed comoving length scale.
A resolution of this problem and definitive results should come from}
the full 2QZ, which should clearly provide even more impressive constraints 
on fine features in density perturbation statistics,
and on the local cosmological parameters $\Omm,$ $\Omega_\Lambda$ and $\wQ$.
}

\maketitle

\dodouble \clearpage :::: 


\def\tdefnsample{
\begin{table}
\caption{Angular subsamples of 
the 2QZ-10K survey. Listed are limits in right ascension 
($\alpha_i$) and declination ($\delta_i$) in coordinate degrees; 
right ascension interval size $\theta$ in great circle degrees; 
approximate solid angle d$\omega$ in square degrees;
number of objects $N$; 
and a rough lower bound to the 
solid angular number density $n$ 
in units of (sq.~deg.)$^{-1}$. Note that the survey is not uniformly
complete within these boundaries; completeness is modelled in the
analysis by use of the angular positions of the observed quasars.
This is why the number density estimates are only lower estimates.
\label{t-defnsample}}
$$\begin{array}{l cccc ccc c} 
\hline
\#: & \alpha_1  & \alpha_2   & \delta_1 &  \delta_2 
& \theta  
& {\mbox d}\omega
& N 
& n \\
1 &    147 & 155 & -3 & 3 
&  8.0 
& 48.0
& 281 
&   5.9\\
2  &    203 & 216 & -3 & 1  
&  13.0 
& 52.0
&  323 
&   6.2\\
3 &    -35 & -30 & -33 & -28   
&   5.0 
& 21.6
& 194 
&   9.0\\
4 &    -17 & -2.5 & -33 & -27 
&  14.5 
& 75.3
& 681 
&   9.0\\
5 &   21 & 31 & -33 & -27 
&10.0 
& 51.9
&  364 
&   7.0\\
6 &    40 & 50 & -33 & -27 
&10.0 
& 51.9
&  535 
&   10.3\\
\mbox{total} &    &  &  & 
&
& 300.7
& 2378
&    \\
\hline
\end{array}$$
\end{table}
}  

\def\fcorrnlowz{ 
\begin{figure} 
{\epsfxsize=86mm 
\zzz{\epsfbox[37 33 548 534]{"`gunzip -c xiz0.6.ps.gz"} } 
{\epsfbox[37 33 548 534]{"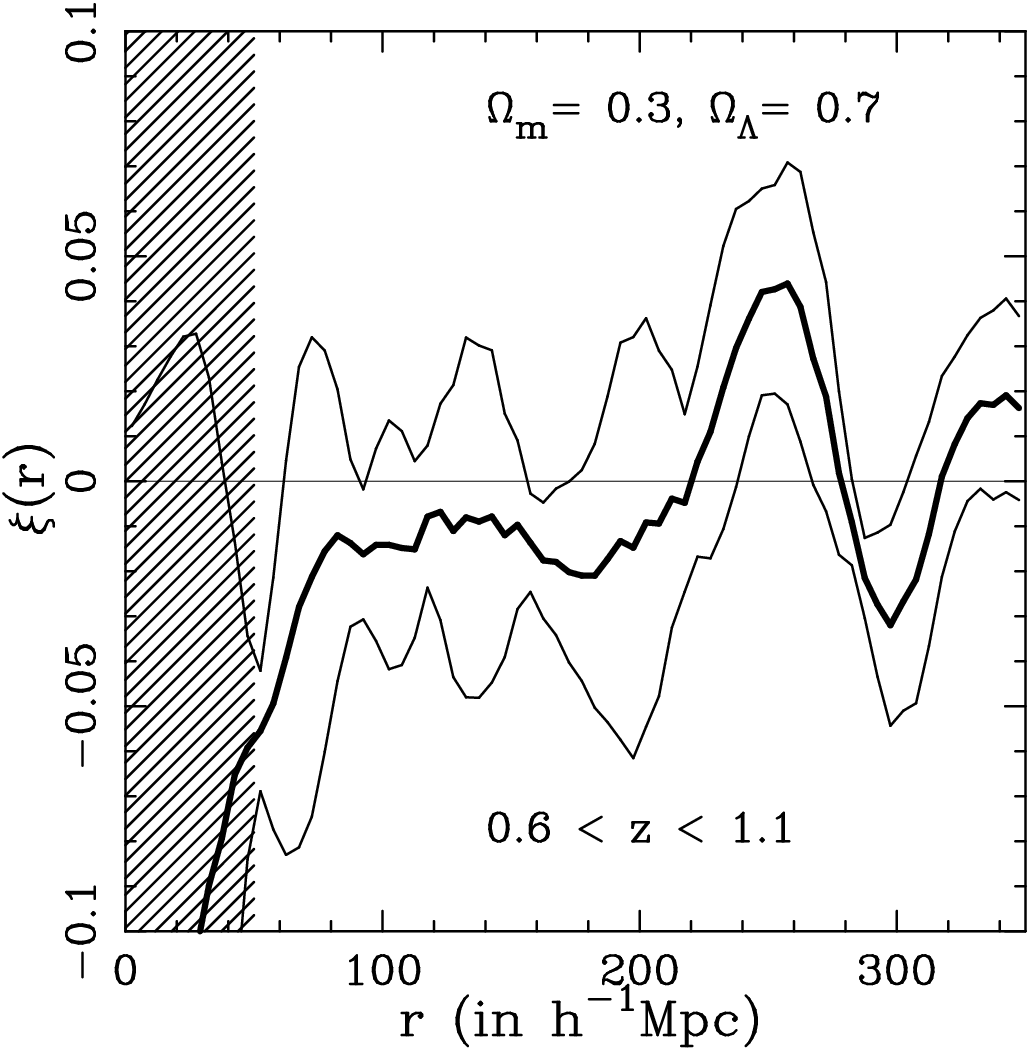"}}{}  } 
\caption[]{
The low-$z$ ($0.6 < z < 1.1$) 
spatial two-point auto-correlation function 
$\xi(r)$, \protect\citep{GroP77}, for separations $r$ in comoving units,  
for a flat universe with ($\Omm=0.3,\Omega_\Lambda=0.7$). 
The $\sigma(r)=15${\hMpc} Gaussian
smoothed mean $\left<\xi\right>$ 
and error bars corresponding to the standard error in the mean 
$\sigma_{\left<\xi\right>}$
are shown by the thick and thin solid lines respectively. 
The small scale $r \ltapprox 25$--$50${\hMpc} correlation is underestimated
(hence the shading) due to the effect of $z$-scrambling in a survey of
narrow beam width containing filaments (see \SS\protect\ref{s-zscramprob} 
and Fig.~\protect\ref{f-zscramprob} for details).
} 
\label{f-corrnlowz} 
\end{figure} 
} 

\def\fcorrnmedz{ 
\begin{figure} 
{\epsfxsize=86mm 
\zzz{\epsfbox[37 33 548 534]{"`gunzip -c xiz1.1.ps.gz"} } 
{\epsfbox[37 33 548 534]{"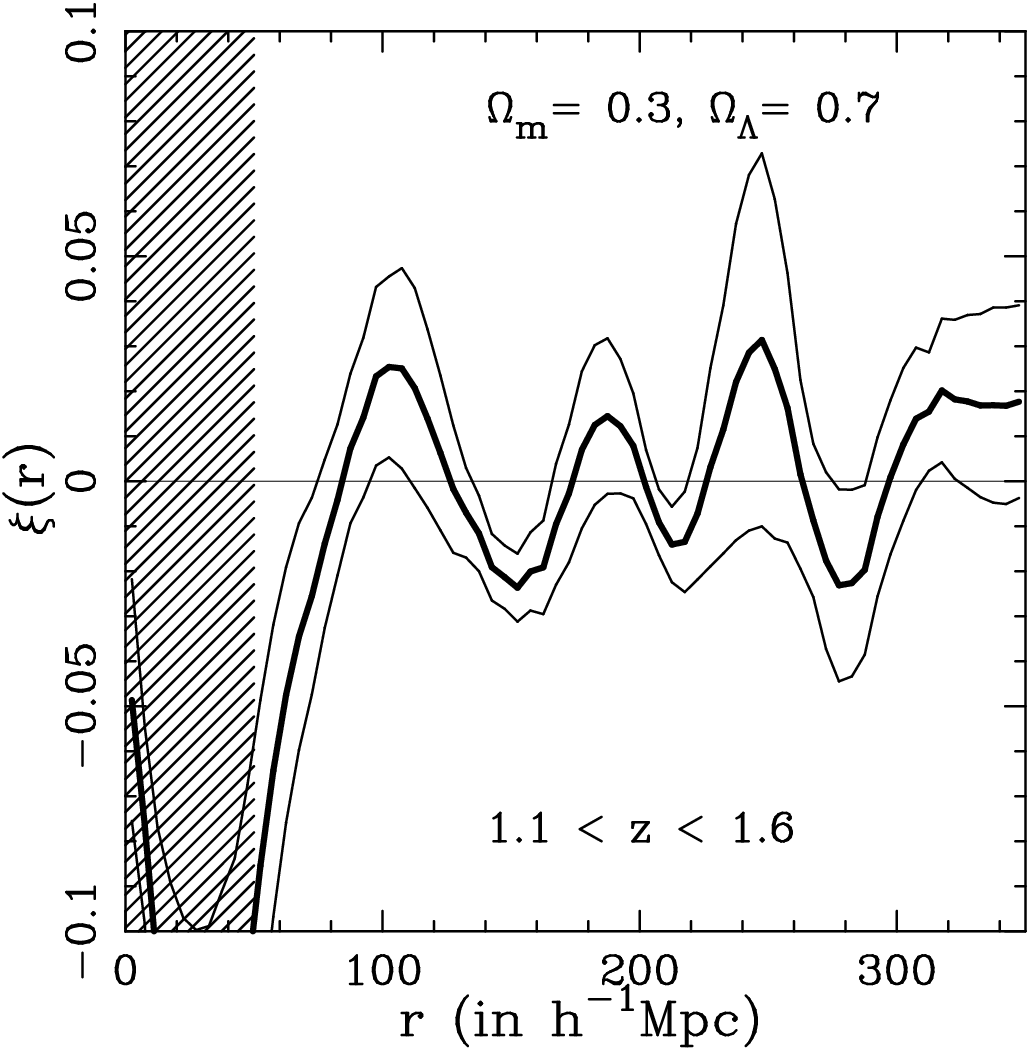"}}{}  } 
\caption[]{
The med-$z$ ($1.1 < z < 1.6$) 
spatial two-point auto-correlation function 
$\xi(r)$, \protect\citep{GroP77}, for separations $r$ in comoving units,  
for a flat universe with ($\Omm=0.3,\Omega_\Lambda=0.7$), as
for Fig.~\protect\ref{f-corrnlowz}. 
} 
\label{f-corrnmedz} 
\end{figure} 
} 

\def\fcorrnhiz{ 
\begin{figure} 
{\epsfxsize=86mm 
\zzz{\epsfbox[37 33 548 534]{"`gunzip -c xiz1.6.ps.gz"} } 
{\epsfbox[37 33 548 534]{"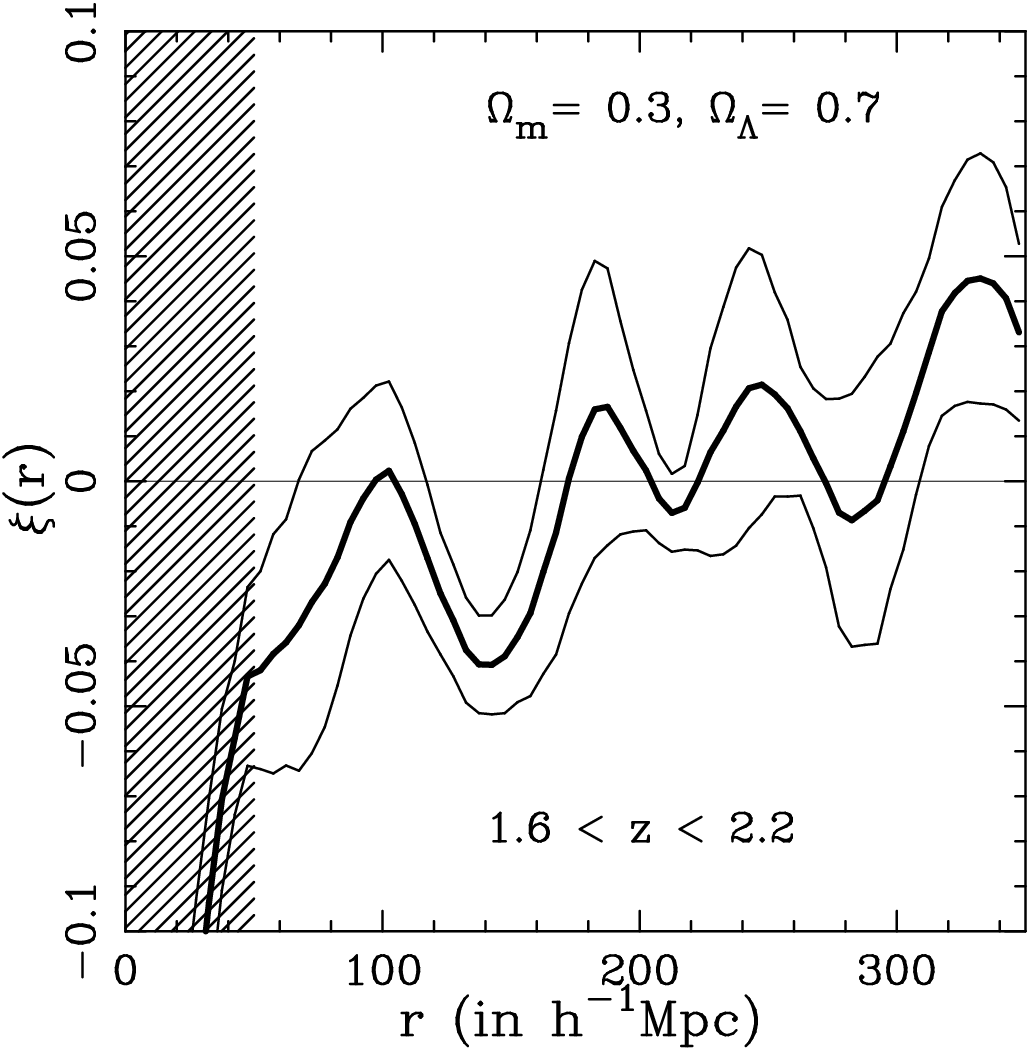"}}{}  } 
\caption[]{
The hi-$z$ ($1.6 < z < 2.2$) 
spatial two-point auto-correlation function 
$\xi(r)$, \protect\citep{GroP77}, for separations $r$ in comoving units,  
for a flat universe with ($\Omm=0.3,\Omega_\Lambda=0.7$), as
for Fig.~\protect\ref{f-corrnlowz}. 
} 
\label{f-corrnhiz} 
\end{figure} 
} 

\def\fcorrnlowzEdS{ 
\begin{figure} 
{\epsfxsize=86mm 
\zzz{\epsfbox[37 33 548 534]{"`gunzip -c xiz0.6EdS.ps.gz"} } 
{\epsfbox[37 33 548 534]{"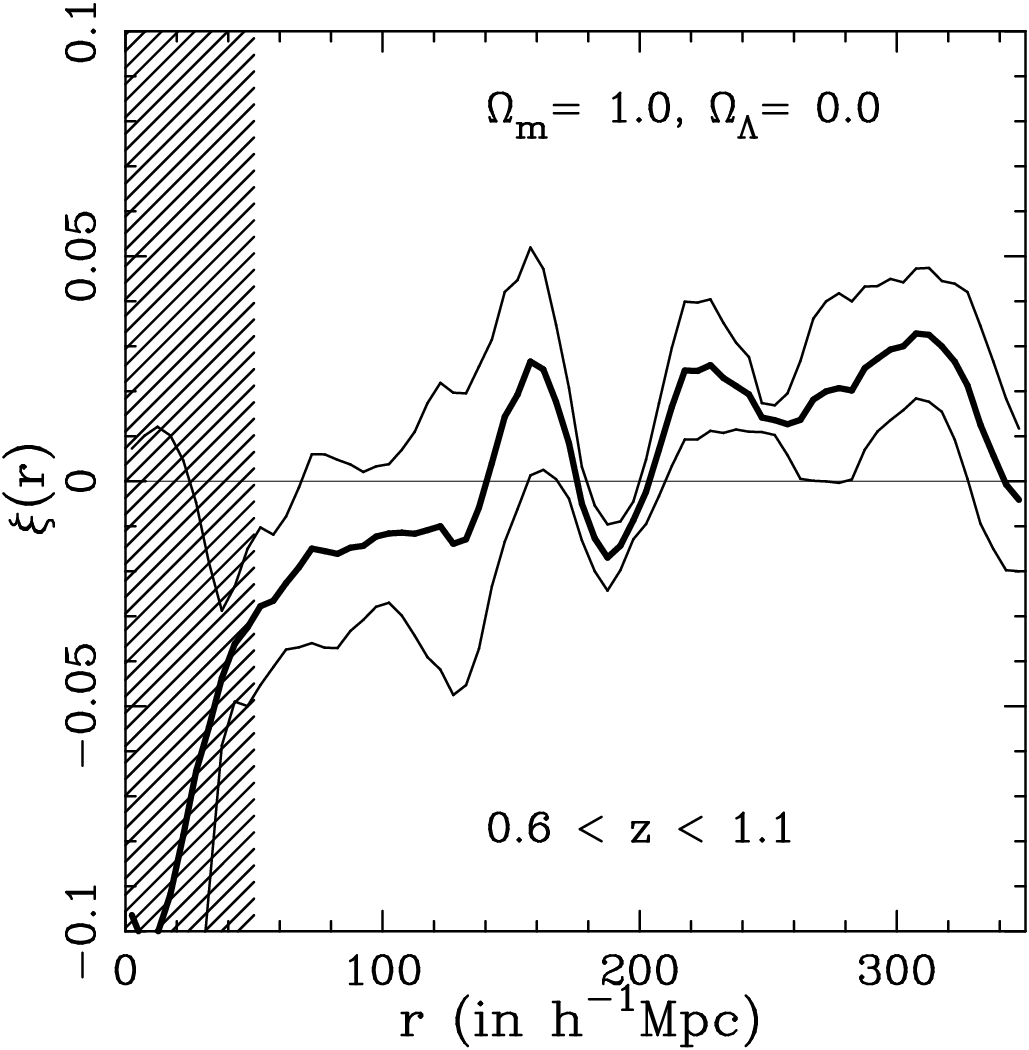"}}{}  } 
\caption[]{
The low-$z$ ($0.6 < z < 1.1$) 
spatial two-point auto-correlation function 
$\xi(r)$, \protect\citep{GroP77}, for separations $r$ in comoving units,  
for a flat universe with ($\Omm=1.0,\Omega_\Lambda=0.0$), as
for Fig.~\protect\ref{f-corrnlowz}. 
} 
\label{f-corrnlowzEdS} 
\end{figure} 
} 

\def\fcorrnmedzEdS{ 
\begin{figure} 
{\epsfxsize=86mm 
\zzz{\epsfbox[37 33 548 534]{"`gunzip -c xiz1.1EdS.ps.gz"} } 
{\epsfbox[37 33 548 534]{"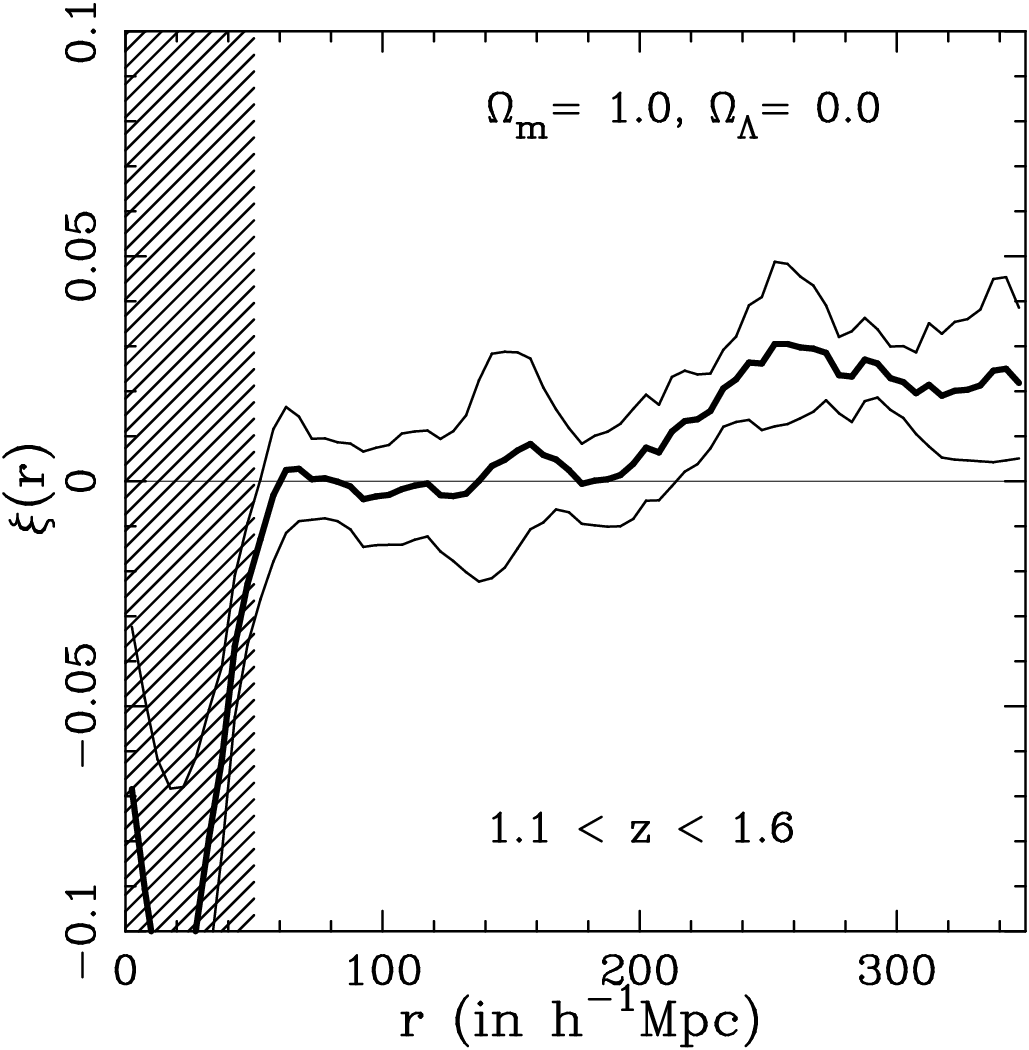"}}{}  } 
\caption[]{
The med-$z$ ($1.1 < z < 1.6$) 
spatial two-point auto-correlation function 
$\xi(r)$, \protect\citep{GroP77}, for separations $r$ in comoving units,  
for a flat universe with ($\Omm=1.0,\Omega_\Lambda=0.0$), as
for Fig.~\protect\ref{f-corrnlowz}. 
} 
\label{f-corrnmedzEdS} 
\end{figure} 
} 

\def\fcorrnhizEdS{ 
\begin{figure} 
{\epsfxsize=86mm 
\zzz{\epsfbox[37 33 548 534]{"`gunzip -c xiz1.6EdS.ps.gz"} } 
{\epsfbox[37 33 548 534]{"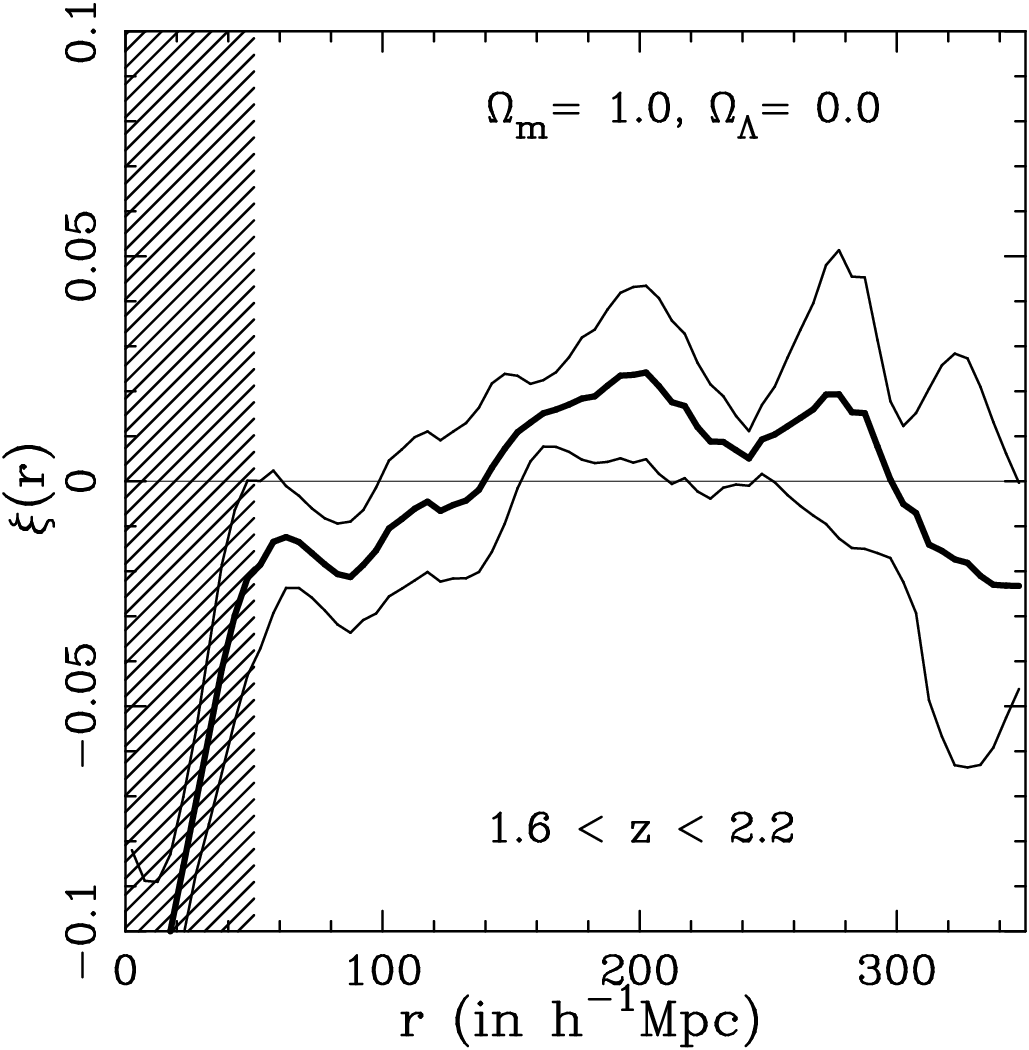"}}{}  } 
\caption[]{
The hi-$z$ ($1.6 < z < 2.2$) 
spatial two-point auto-correlation function 
$\xi(r)$, \protect\citep{GroP77}, for separations $r$ in comoving units,  
for a flat universe with ($\Omm=1.0,\Omega_\Lambda=0.0$), as
for Fig.~\protect\ref{f-corrnlowz}. 
} 
\label{f-corrnhizEdS} 
\end{figure} 
} 

\def\fcorrnlowzL{ 
\begin{figure} 
{\epsfxsize=86mm 
\zzz{\epsfbox[37 33 548 534]{"`gunzip -c xiz0.6L.55.ps.gz"} } 
{\epsfbox[37 33 548 534]{"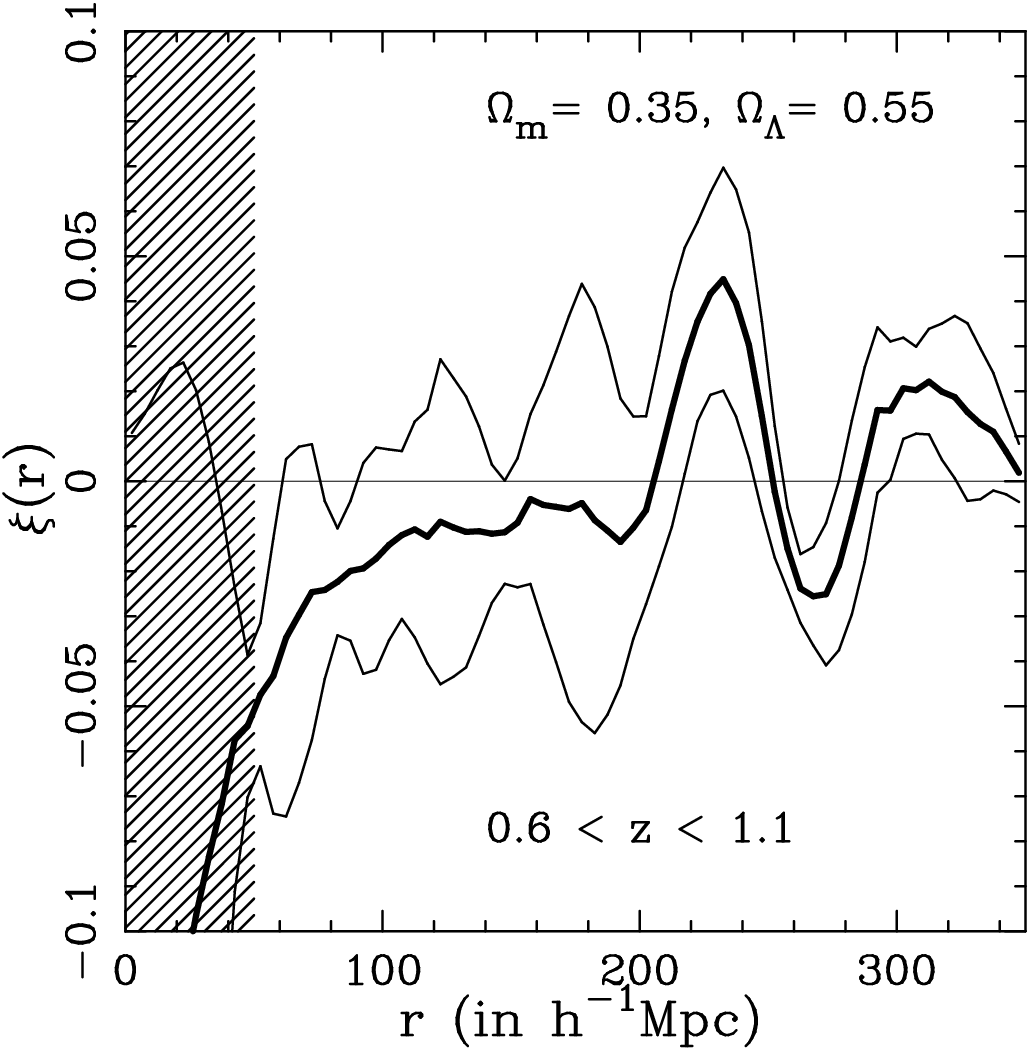"}}{}  } 
\caption[]{
The low-$z$ ($0.6 < z < 1.1$) 
spatial two-point auto-correlation function 
$\xi(r)$, 
for an hyperbolic universe with ($\Omm=0.35,\Omega_\Lambda=0.55$), as
for Fig.~\protect\ref{f-corrnlowz}. 
} 
\label{f-corrnlowzL} 
\end{figure} 
} 

\def\fzscramprob{ 
\begin{figure} 
{\epsfxsize=86mm 
\zzz{\epsfbox[59 40 531 308]{"`gunzip -c zscram.ps.gz"} } 
{\epsfbox[59 40 531 308]{"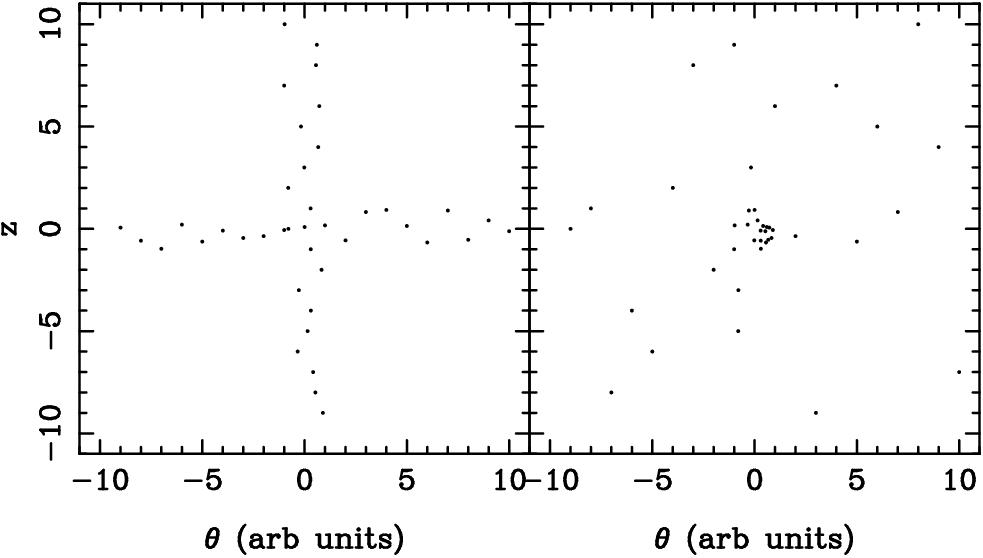"}}{}  } 
\caption[]{
Toy model showing why small scale correlations can be
underestimated by $z$-scrambling when correlated structures 
exist on the scale of the survey volume and parallel to 
the $\theta$ and $z$ directions (\SS\protect\ref{s-zscramprob}). 
The left-hand panel 
shows a toy model with filaments, 
parallel to these two directions,
and each composed of 20 points.
The right-hand panel shows a $z$-scramble of the distribution
in the left-hand panel. The two ``favoured'' 
values of $\theta\approx 0 $ and $z\approx0$ in the original filaments
(length units are 
arbitrary in the two directions) result in many $z$-scrambled 
pairs containing both these ``favoured'' values. Hence, many more
{\em close} pairs exist in the $z$-scrambled distribution than
in the original, correlated distribution.  So, relative to the 
$z$-scrambled distribution, the original distribution is anti-correlated
for pair separations up to the scale of the filament thickness, 
leading to underestimates of $\xi$ on these small scales.
} 
\label{f-zscramprob} 
\end{figure} 
} 

\def\fABC{ 
\begin{figure} 
{\epsfxsize=86mm 
\zzz{\epsfbox[29 26 459 488]{"`gunzip -c stat01_100.ps.gz"} } 
{\epsfbox[29 26 459 488]{"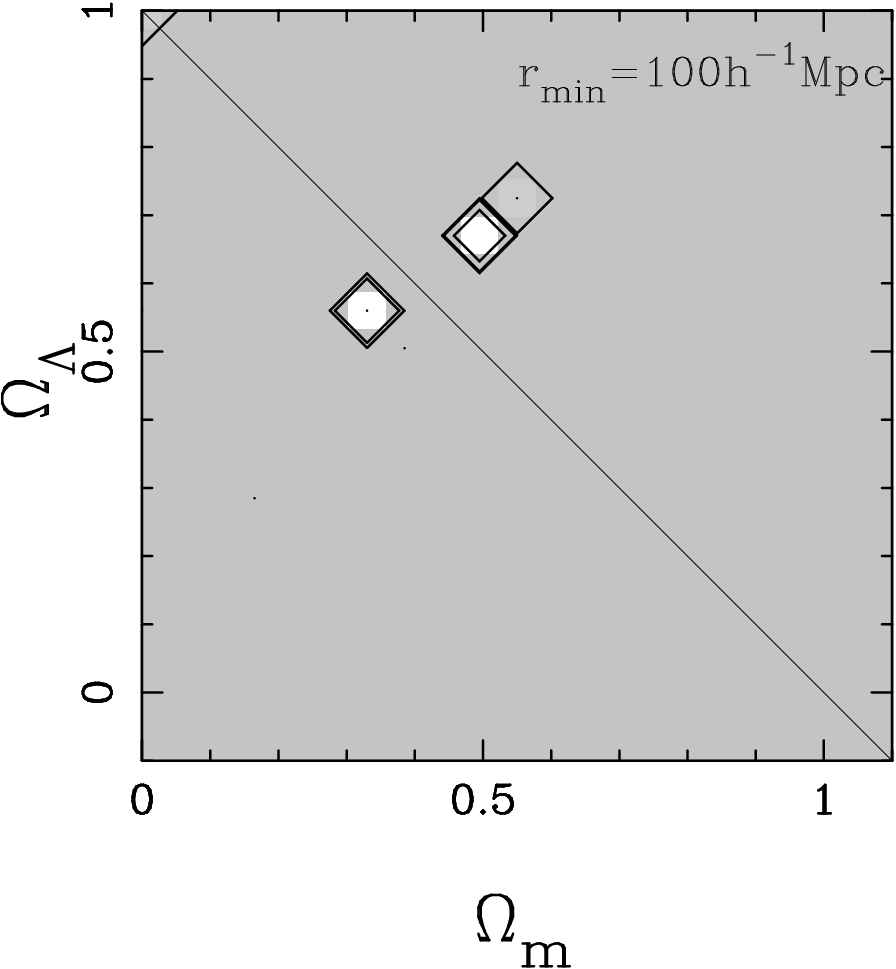"}}{}  } 
\caption[]{Confidence intervals for rejecting the hypothesis that the 
first local maximum in the spatial 
correlation function at $r > 100${\hMpc} occurs at the same comoving
position in the low-$z$, med-$z$ and hi-$z$ redshift intervals of 
the 2QZ-10K,
for various hypotheses on the values of $(\Omm,\Omega_\Lambda)$. 
The uncertainty in the estimate of 
the position of a local maximum is $\Delta\llss=5${\hMpc}.
Rejection levels of
$68\% >1-P$, $95\% > 1-P$,
$99.7\% > 1-P$ and
$1-P >99.7$ are indicated in this figure and
in Figs~\protect\ref{f-ABCb},\protect\ref{f-ABCc}, 
with gray shadings from white to gray as
a visual aid. In this figure,
the two white squares indicate $(\Omm,\Omega_\Lambda)$ values 
satisfying $68\% >1-P$, and nearly all of the figure lies in 
the $1-P >99.7$ domain. Due to stronger signals 
in Figs~\protect\ref{f-ABCb},\protect\ref{f-ABCc}, 
the contours in these figures are better labelled.
A thin line indicating a flat universe ($\Omm+\Omega_\Lambda=1$) is
indicated in this and the following figures.
}
\label{f-ABC}
\end{figure}
}

\def\fABCb{ 
\begin{figure} 
{\epsfxsize=86mm 
\zzz{\epsfbox[29 26 459 488]{"`gunzip -c stat01_150.ps.gz"} } 
{\epsfbox[29 26 459 488]{"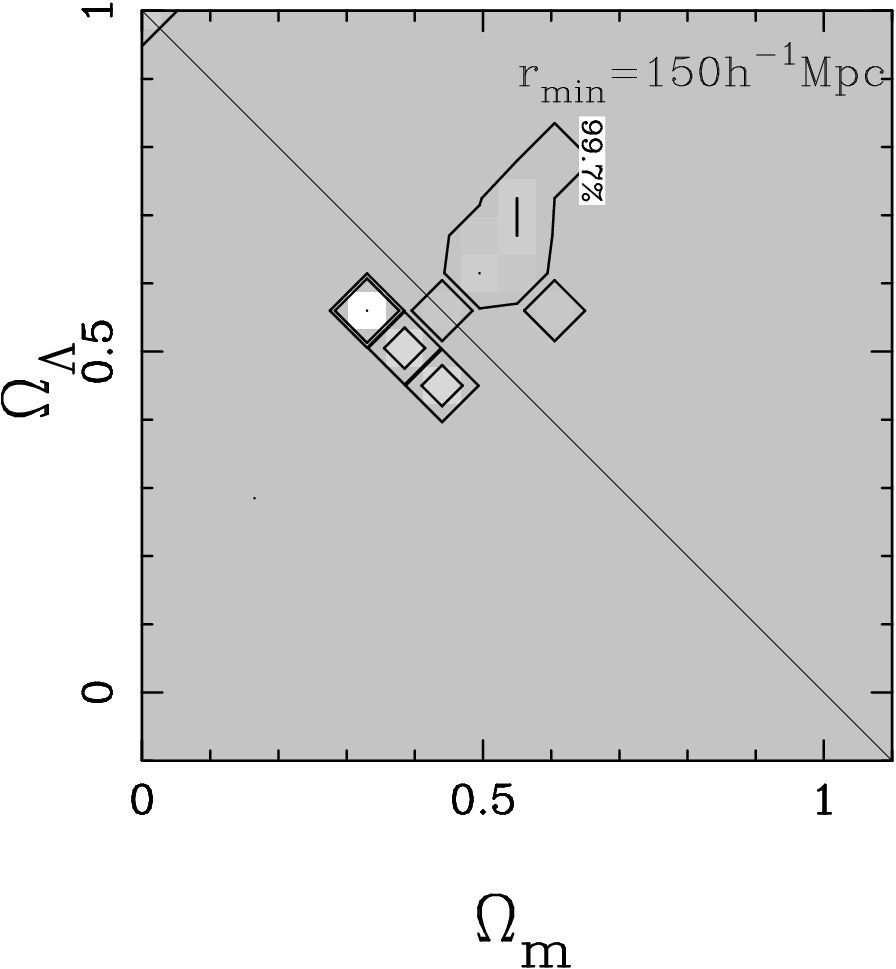"}}{}  } 
\caption[]{As for Fig.~\protect\ref{f-ABC}, 
confidence intervals for rejecting the hypothesis that the 
first local maximum in the spatial 
correlation function at $r > 150${\hMpc} in all three redshift
intervals.
}
\label{f-ABCb}
\end{figure}
}

\def\fABCc{ 
\begin{figure} 
{\epsfxsize=86mm 
\zzz{\epsfbox[29 26 459 488]{"`gunzip -c stat01_200.ps.gz"} } 
{\epsfbox[29 26 459 488]{"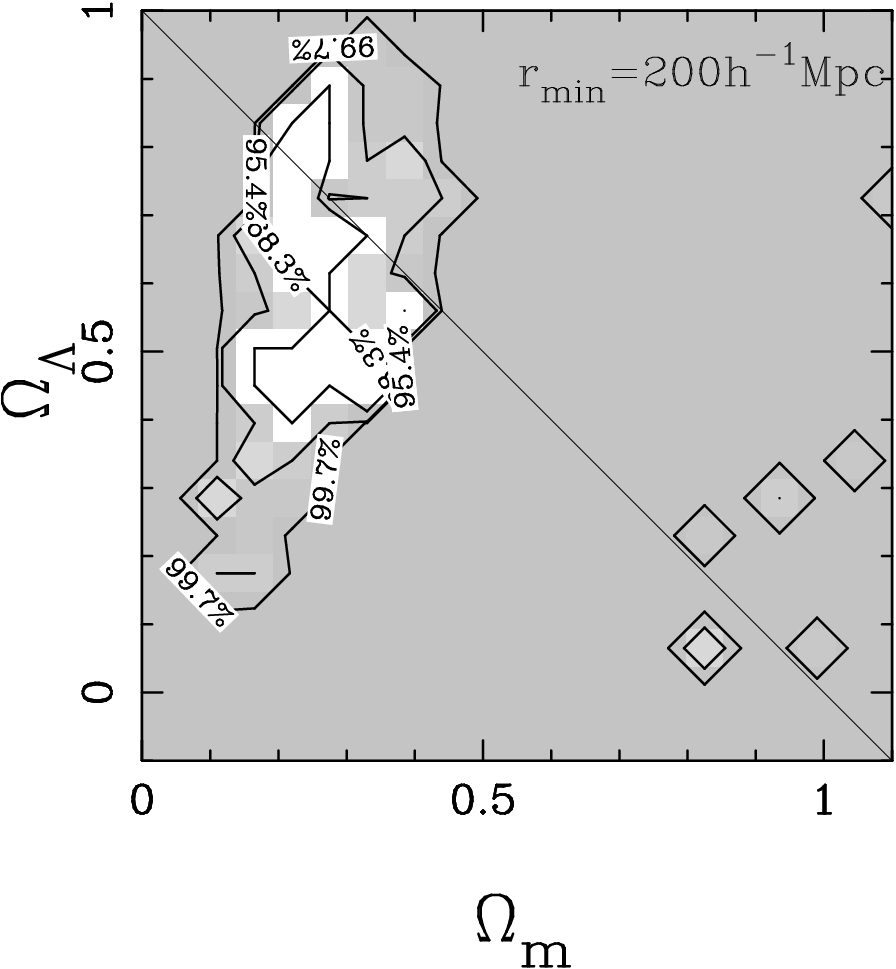"}}{}  } 
\caption[]{As for Fig.~\protect\ref{f-ABC}, 
confidence intervals for rejecting the hypothesis that the 
first local maximum in the spatial 
correlation function at $r > 200${\hMpc} in all three redshift
intervals.
}
\label{f-ABCc}
\end{figure}
}

\def\fABCwq{ 
\begin{figure} 
{\epsfxsize=86mm 
\zzz{\epsfbox[29 26 459 488]{"`gunzip -c stat01_200_w.ps.gz"} } 
{\epsfbox[29 26 459 488]{"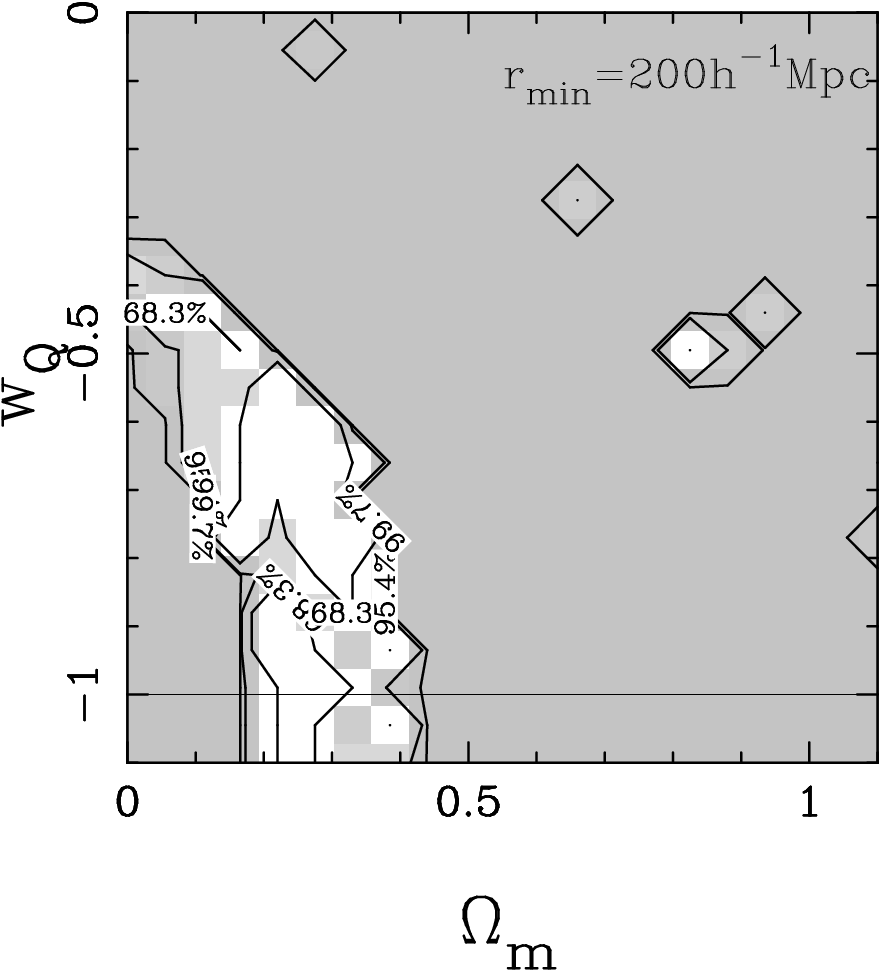"}}{}  } 
\caption[]{As for Fig.~\protect\ref{f-ABC}, 
confidence intervals for rejecting the hypothesis of a local correlation 
function maximum 
at $r > 200${\hMpc} present in all three 2QZ-10K redshift intervals,
for a flat, effective quintessence model, for various hypotheses
on the values of $(\Omm,\wQ)$. A thin line where $\wQ\equiv 1$, i.e. 
equivalent to a cosmological constant model, is indicated in the figure.
}
\label{f-ABCwq}
\end{figure}
}

\def\fD{ 
\begin{figure} 
{\epsfxsize=86mm 
\zzz{\epsfbox[29 26 459 488]{"`gunzip -c stat02.ps.gz"} } 
{\epsfbox[29 26 459 488]{"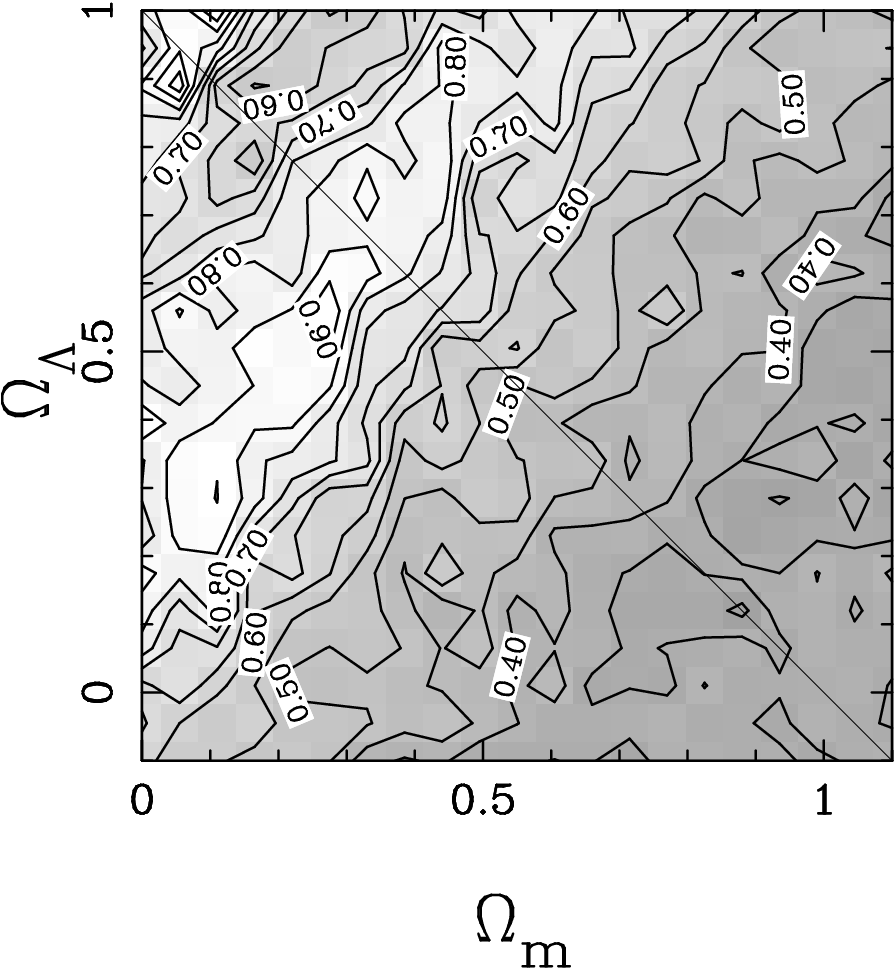"}}{}  } 
\caption[]{Values of the difference statistic $D$
[Eq.~(\protect\ref{e-D})], representing the average absolute slope
of $\xi(r)$ in the interval $200${\hMpc} $\le r \le 300${\hMpc}
over all three redshift intervals,
which would depend monotonically on $\Omm$ and $\Omega_\Lambda$ if
fluctuations in this interval were due only to noise and not to
a genuine cosmological signal. The maximum value of $D$ is $D=1$.
Contours from $D=0.4$ to $D=0.9$, in intervals of $\Delta D=0.05$
are indicated. Shading is darker for lower values of $D$. 
It is clear that $D$ does not increase monotonically from
$(\Omm=1.1,\Omega_\Lambda=-0.1)$ to $(\Omm=0.0,\Omega_\Lambda=1.0)$.
The excess values of $D$, ie. the excess 
average absolute differences, with respect to 
the underlying trend, occur in a region peaked 
at $(\Omm\approx 0.25, \Omega_\Lambda\approx0.55$).
}
\label{f-D}
\end{figure}
}

\def\fDp{ 
\begin{figure} 
{\epsfxsize=86mm 
\zzz{\epsfbox[29 26 459 488]{"`gunzip -c stat02poiss.ps.gz"} } 
{\epsfbox[29 26 459 488]{"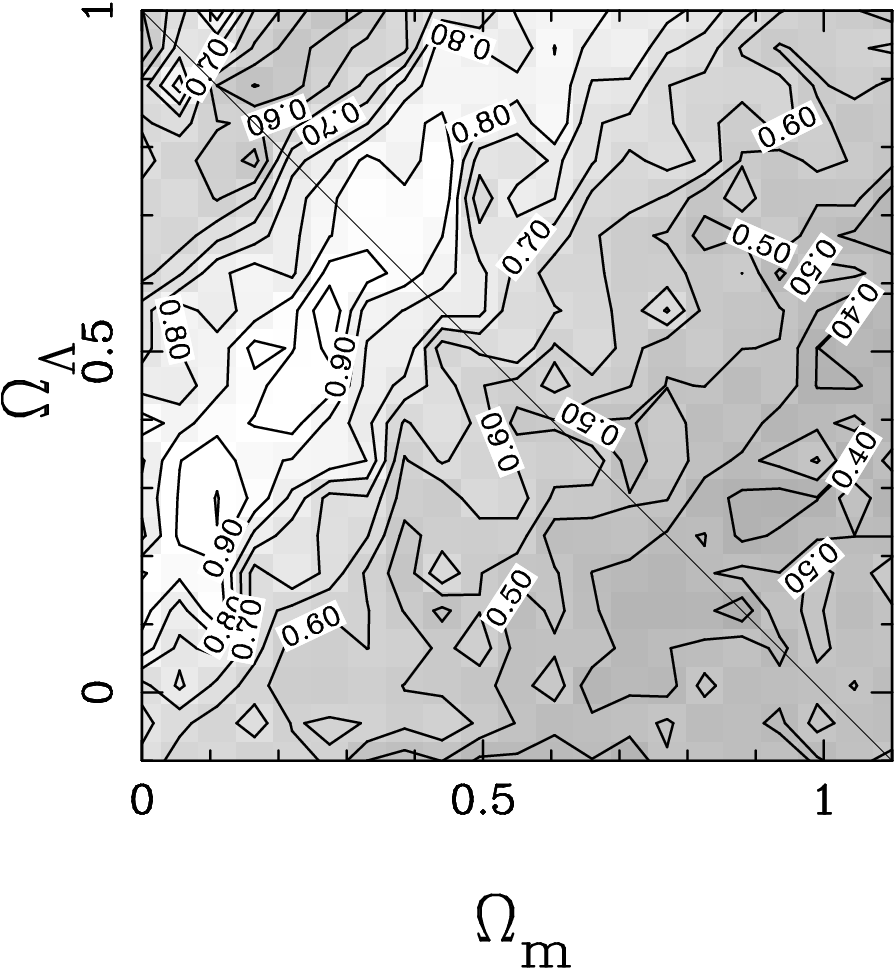"}}{}  } 
\caption[]{Values of the Poisson-corrected difference statistic $D'$
[Eq.~(\protect\ref{e-Dp})], as for Fig.~\protect\ref{f-D}, but where $D$
is approximately  
corrected for the monotonic 
variation as a function of expected Poisson noise, yielding $D'$.
It is clear that excess values of $D'$ occur 
near $(\Omm\approx 0.25, \Omega_\Lambda\approx0.55$) and pass through
a linear degeneracy from this point to $(\Omm\approx 0.1, 
\Omega_\Lambda\approx 0.3)$. This confirms that a non-Poisson signal
is present in $\xi(r)$ for a pair $(\Omm, \Omega_\Lambda)$ lying
somewhere in this degeneracy region.
}
\label{f-Dp}
\end{figure}
} 

\def\fDphun{ 
\begin{figure} 
{\epsfxsize=86mm 
\zzz{\epsfbox[29 26 459 488]{"`gunzip -c stat02poiss100.ps.gz"} } 
{\epsfbox[29 26 459 488]{"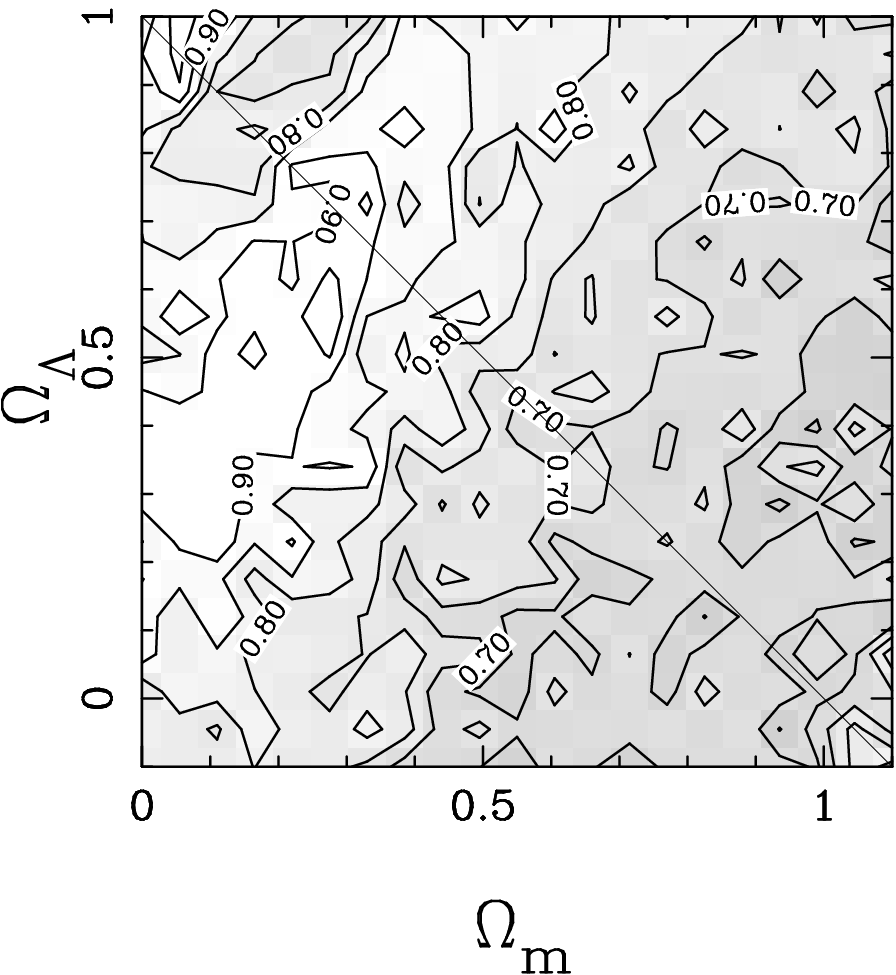"}}{}  } 
\caption[]{Values of the Poisson-corrected difference statistic $D'$,
as for Fig.~\protect\ref{f-Dp}, but for the interval 
$100${\hMpc}~$\le r \le 350${\hMpc}. The 
excess values of $D'$ seen in Fig.~\protect\ref{f-Dp} 
near $(\Omm\approx 0.25, \Omega_\Lambda\approx0.55$) remain present.
Imperfect correction for the Poisson error also yields a background
which is not quite flat, so that high $D'$ values also appear near
$\Omm\approx 0.05,\Omega_\Lambda\gtapprox 0.95$, but are due to 
Poisson error and not an excess relative to Poisson error.
}
\label{f-Dp100}
\end{figure}
} 


\section{Introduction} \label{s-intro}

Increasing observational evidence is mounting in favour of 
fine structure (local maxima) in the power spectrum or
correlation function of the spatial distribution of extragalactic
objects, at scales $\sim100$--$200${\hMpc}.
At redshifts $z \sim 2-3$, these features have been found 
among quasars and Lyman break galaxies 
\citep{Deng94,BJ99,RM00,RM01},
and among low redshift objects, 
similar features have been found by numerous groups  
(e.g. 
\citealt*{Bro90,Bro99,daCosta92,daCosta93,BauE93,BauE94,GazB98,Einasto94,Einasto97corr,Einasto97nat,Deng96,Tago01}; 
see \citealt{Guzzo99} for a recent review, 
or \citealt{KirCh00} for a very extensive reference list).

Because the scale $\sim100$--$200${\hMpc} is well into the linear regime
of density perturbations and well above the present-day turnaround
scale at which matter has had the time to collapse by self-gravity,
{\em these features should occur at fixed comoving length scales},
provided that the correct values of the local cosmological parameters
$\Omm$ (the matter density parameter), $\Omega_\Lambda$ (the dimensionless
cosmological constant) and $\wQ$ (the effective quintessence parameter,
e.g. \citealt{Efst99}) are used to convert redshifts and angular
positions to three-dimensional positions in 
three-dimensional (proper distance, comoving) 
space of the appropriate curvature 
(e.g. 
\citealt{Rouk01arc}).

At low redshifts ($z \ltapprox 0.1$), peculiar velocities and the use of 
incorrect values of the local cosmological parameters make 
claims either for or against these 
fine features difficult. On the other hand, 
at high redshifts ($1\ltapprox z \ltapprox 3$), while peculiar velocities
have a much smaller effect and assumptions on the local cosmological
parameters are more often stated clearly than for low redshift analyses,
the main difficulty is 
the sparsity of observational data in homogeneous and large volume 
catalogues.

In \citet{RM00,RM01}, the objective prism survey of quasar 
candidates by \citet*{IovCS96} was analysed, using the {\em a priori} 
claim, primarily from the low redshift analyses, that a local 
maximum in the power spectrum or the spatial correlation function
exists at  $k = 2\pi/\llss$, 
where $\llss=130\pm10${\hMpc}. In a fourier analysis of the initial
release of data from the 2dF Quasar Survey (hereafter, 2QZ-10K, 
\citealt{Croom01}),
\citet{Hoyle01} claim marginal evidence for the existence of a ``spike''
at 65--90{\hMpc}, depending on the values of the local cosmological
parameters.

Spectroscopic followup of a small part of the \citet{IovCS96} 
survey (P.~Petitjean, private
communication) suggests that it suffers from 
contamination by stars and by quasars at redshifts other than
those estimated by \citet{IovCS96}, so it is possible that the 
features found were either underestimated or overestimated in strength
and/or shifted in length scale with respect to the true signal.

On the contrary, the 2QZ-10K consists uniquely of diffraction 
grating spectroscopic
redshifts, so should be free of this contamination and potentially
provide a cleaner signal.

However, extreme care is required in the analysis of 
the 2QZ-10K survey, 
because of:
\begin{list}{(\alph{enumi})}{\usecounter{enumi}}
\item the ``angular selection function'', i.e. 
the very anisotropic distribution of this initial 
release within the two survey regions (e.g. fig.~1, \citealt{Croom01});
\item the two sources of incompleteness in the regions observed so far:
``spectroscopic incompleteness'' (fraction of objects observed but with
unsuccessful spectroscopic identification) and
``coverage incompleteness'' (fraction of target objects in a given field
not yet spectroscopically observed); and
\item the initial selection of the target catalogue using 
$ub_{\mbox{\small J}}r$ multi-colour photometry using photographic plates.
\end{list}

In this paper, a spatial correlation function analysis of the 2QZ-10K
is performed, as in \citet{RM01},
in order to minimise spurious artefacts from factors (a)--(c) and
enable a ``comoving standard ruler'' analysis of the data. 
Moreover, in order to avoid the {\em a priori} assumption
of the scale at which feature(s) defining the standard ruler should occur,
the constraint is relaxed to only 
require {\em consistency} of the scales of 
feature(s) between different redshift intervals of the 2QZ-10K, 
without assuming the scale(s) at which these feature(s) occur.

As will be seen below, consistency of length scales in 
the comoving spatial correlation function between different redshifts
implies constraints on the local cosmological parameters {\em independently
of cosmic microwave background, type Ia supernovae, or other constraints on 
these parameters}.

The observational catalogue and method of analysing it 
in a way that deals with points (a)--(c) above
are briefly described in 
\SS\ref{s-ob+meth}, results are presented in \SS\ref{s-results},
the question of whether or not the comoving local maximum detected
is a genuine, cosmological feature rather than just a noise 
fluctuation
is discussed in \SS\ref{s-discuss} 
and conclusions are presented in \SS\ref{s-conclu}.

An almost (i.e. perturbed) 
Friedmann-Lema\^{\i}tre-Robertson-Walker cosmological model
is assumed here, optionally including a quintessence relation. 
The Hubble constant is parametrised 
as $h\equiv H_0/100\,$km~s$^{-1}$~Mpc$^{-1}.$ Comoving coordinates are
used throughout 
[i.e. `proper distances', eq.~(14.2.21), \citet{Wein72}, 
equivalent to `conformal time'
if $c=1$; see \citet{Rouk01arc} for a self-contained and 
conceptually simple presentation of how to calculate three-dimensional
comoving separations for a curved space, by considering these as arc-lengths
in four-dimensional space].
Values of the density parameter, $\Omm$, and 
the dimensionless cosmological constant,
$\Omega_\lambda$, are indicated where used.

For convenience, the reader is reminded here of the
dependence of proper distance on $\Omm$ and $\Omega_\Lambda$, i.e.
 \begin{equation}
d(z) 
= {c \over H_0} \int_{1/(1+z)}^1 
{ \mbox{\rm d}a \over a \sqrt{\Omm /a - \Omega_\kappa + \Omega_\Lambda a^2} },
\label{e-defdprop}
\end{equation}
where $a$ is the scale factor, 
\begin{equation}
\Omega_\kappa \equiv \Omm + \Omega_\Lambda -1
\label{e-defcurv}
\end{equation}
is the (dimensionless) curvature of the observational Universe and 
\begin{equation}
R_C \equiv {c \over H_0} { 1 \over \sqrt{ | \Omega_\kappa| } }
\end{equation}
is its curvature radius.

Effective quintessence distance-redshift relations 
in a flat universe are also considered, where 
$\Omega_Q \equiv 1- \Omm$ and 
$\Omega_Q a^{-3(1+\wQ)}$ is substituted for
$\Omega_\Lambda$ in Eq.~(\ref{e-defdprop}), 
where 
$\wQ$ is considered to be a constant (e.g. \citealt{Efst99}).
The value $\wQ=-1$ corresponds to a standard cosmological
constant metric.


\section{Catalogue and analysis} \label{s-ob+meth}

\subsection{The 2QZ} \label{s-choosecat}

The 2QZ-10K is described in \citet{Croom01}. 
This release only includes quasars in regions of 85\% or greater 
``spectroscopic completeness''.

In order to minimise noise generated by ``coverage incompleteness'',
only those quasars observed in regions with at least 80\% coverage are
included in the present analysis. The six highest solid angular density 
regions are chosen from fig.~1 of \citet{Croom01} and listed 
in Table~\ref{t-defnsample}, providing six independent sub-samples
from which to obtain estimates of $\xi(r)$.

Together, these completeness criteria 
imply a minimum of about 70\% total completeness, 
apart from the incompleteness implied by initial selection of target
objects using photographic $ub_{\mbox{\small J}}r$ photometry
(as opposed to spectroscopic observation of {\em all} objects, selected
in a single filter from CCD photometry).

In \citet{RM00,RM01}, the high number density, objective prism 
survey near the South Galactic Pole \citep{IovCS96} was
analysed. The right ascension and declination subsamples of the
survey in the \citet{RM01} analysis 
had solid angular number densities of $9.8$ and $8.8$ quasars/sq.deg{.}
respectively.

\tdefnsample

The regions of the 2QZ so far completed have solid angular 
number densities similar to this
(lower bounds are listed in Table~\ref{t-defnsample}), 
but since a larger redshift interval is covered, the comoving
spatial densities are about a factor of three lower.
The signal-to-noise ratios 
of a peak (local maximum) in $\xi(r)$ 
may not be as high in the 2QZ as in the \citet{IovCS96} survey.
Note that the two surveys are
independent in angular and redshift range, 
apart from a slight overlap.


\subsection{Method} \label{s-method}

\subsubsection{Calculation of correlation functions}
The correlation functions are calculated in three-dimensional 
curved space via 
$\xi(r) = (DD- 2DR/n +RR/n^2)/(RR/n^2)$ 
where $DD$, $DR$ and $RR$ indicate numbers of data-data, data-random 
and random-random quasar pairs respectively \protect\citep{LS93}, 
and $n=20$ times more 
random points than data points are used. 
The random catalogues use (i) the exact set of angular positions
of the catalogue quasars, and (ii) random permutations 
(``$z$ scrambles'', {\SS}IIIb in \protect\citealt{Osmer81}) 
of the observational set of redshifts. 
Bin size is $\Delta r = 5${\hMpc}.

In each of the approximately equal redshift intervals 
$0.6 < z < 1.1$ (``low-$z$''),
$1.1 < z < 1.6$ (``med-$z$'') and
$1.6 < z < 2.2$ (``hi-$z$''), 
$\xi(r)$ is calculated for each of the six angular sub-samples 
defined in Table~\ref{t-defnsample}.

Within each redshift interval,
mean values of $\xi(r)$ and standard errors in the mean of $\xi(r)$ 
across the six sub-samples are calculated, since these are independent
sub-samples.
Each function $\xi$ is smoothed by a Gaussian with $\sigma=15${\hMpc}. 
Weighting in inverse proportion to the number of pairs 
in each angular sub-sample 
is applied in order to obtain signal in proportion to 
the respective numbers of pairs. 

\subsubsection{Consistency of spikes at all redshifts}

Although $\llss\approx 130\pm10${\hMpc} has been claimed by several
authors to be the scale of a local maximum in the power spectrum
or correlation function, this is {\em not assumed} here.

Spikes are detected here as the first local maximum in $\xi(r)$ 
above a given minimum separation $\rmin$, where 
$\rmin=100, 150$ or $200${\hMpc}. 
In order for the maxima not to be {\em too} local (e.g. a slightly 
high value in one bin at $r$ 
surrounded by slightly lower values in immediately 
adjacent bins at $r-\Delta r$ and $r+\Delta r$), 
a smoothing by a Gaussian of width $\sigma=15${\hMpc} is applied when
searching for a local maximum.

For a local maximum to be considered detected and useful for constraining
local cosmological parameters, the estimates of its position $\rmax$ 
should be consistent in all three redshift ranges. Quantitatively, this
consistency is defined by requiring the values of $\rmax$ 
to agree within $\Delta r=5${\hMpc} between 
the low-$z$ and the med-$z$ sub-samples,
and within $\Delta r=5${\hMpc} between 
the med-$z$ and the hi-$z$ sub-samples, assuming Gaussian errors. 

So, for a given value of $\rmin$, 
the consistency hypothesis implies rejection probabilities $P$ and
confidence levels $1-P$.

The above test is applied without any criteria regarding the
significance of individual maxima, in order to be as parameter-free as
possible, though at the risk of including noise. 

For the sake of discussion, an estimate of the signal-to-noise ratio
of a local maximum can be quantified as follows.

For a maximum at $\rmax$, define $\xi^r,$ $\xi_-$ and $\xi_+$ as 
the value of $\xi$ at $\rmax$ and the values of $\xi$ 
at the first minima at $r_-$ and $r_+$ below and above 
$\rmax$ respectively, and 
take the root mean square value of $\sigma_{\left<\xi\right>}(r)$ for 
$r\in [r_-,r_+]$. Then,
\begin{equation} 
(S/N) \equiv { \xi^r - (\xi_- + \xi_+)/2 \over 
\sqrt{ \left< \sigma^2_{\left<\xi\right>}(r) \right> }   } 
\label{e-defnsn} 
\end{equation} 
is the definition of signal-to-noise used for comments below.

\subsubsection{Advantages of $z$-scrambling}
\begin{list}{(\roman{enumi})}{\usecounter{enumi}}
\item
 The angular positions are used 
because the angular selection function of the 2QZ is difficult
to quantify by any other method, as stated above (a), in \SS\ref{s-intro}.
\item
 The use of $z$-scrambling \citep{Osmer81} is 
adopted for obtaining the `random' redshift distributions, 
as in \citet{RM01}, 
in order to avoid redshift selection effects, as mentioned
above in (c) of \SS\ref{s-intro} (c.f. \citealt{Scott91}).  
\end{list}

The problems of incompleteness (b) are also minimised by this 
method of generating the random distributions.

\fzscramprob

Note that incompleteness is more likely to lead to noise than to
spurious effects: if a ``void'' traced by quasars has several quasars
missing, then it will be poorly traced, but will not be changed in
diameter.

This method implies conservative results. 
Since some of the real, very large scale (e.g. $\sim1${\hGpc}), 
but small amplitude correlations 
which may occur in the radial direction
occur
in both the real and the so-called random catalogues, these will 
cancel out along with the selection effect correlations that 
are (deliberately) cancelled, 
and be lost from the final signal. 
(These very large scale correlation 
cannot occur in the tangential direction in the 2QZ-10K, 
because of the angular selection function and division into six sub-samples
by angle.)

\fcorrnlowz
\fcorrnmedz
\fcorrnhiz

\fcorrnlowzEdS
\fcorrnmedzEdS
\fcorrnhizEdS

So, this is an advantage in the sense that 
selection effects are cancelled; but can also be seen as a disadvantage
in that the amplitude of real, non-zero values in 
the correlation function, in particular on large scales,
may be underestimated. Since the phenomenon of interest here
is the use of {\em local} feature(s) (first derivatives) 
in the correlation function as a standard ruler for
obtaining geometrical constraints,
the advantage largely outweighs the 
disadvantage: the method is conservative.

While absolute estimates of the correlation function (or power spectrum) 
clearly have cosmological relevance, the optimal methods for correcting
for angular and redshift selection effects are likely to be considerably
different from the method adopted here, and are not considered in this
paper.

\subsubsection{Small $r$ disadvantage of $z$-scrambling}
\label{s-zscramprob}

The $z$-scrambling technique can have a disadvantage if
\begin{list}{(\roman{enumi})}{\usecounter{enumi}}
\item either selection effect correlations or intrinsic correlations
exist on a scale comparable to that of two 
of the dimensions of the survey volume, 
{\em and}
\item these correlations occur as filaments which are 
in some cases aligned with
the tangential ($\theta$) and radial ($z$) directions.
\end{list}

The effect is an overestimate of the numbers of 
close pairs in the random simulations,
on the scale of the ``thickness'' of the ``filaments'', and, hence,
an underestimate of $\xi(r)$ on scales $r$ smaller than the
``filament thickness''.

This is illustrated in a toy model in Fig.~\ref{f-zscramprob}, with
a qualitative explanation in the caption.

An algebraic explanation of the effect 
can be made by writing the original distribution as points $f_1^i$ 
in filament 1 and points $f_2^i$ 
in filament 2, at approximately Cartesian 
coordinates $(\theta,z)$ for $i=1,10$:
\begin{eqnarray}
f_1^i&\approx &(0,\pm i) \nonumber\\
f_2^i&\approx &(\pm i,0) 
\label{e-zscram}
\end{eqnarray}
where ``$\approx$'' indicates that the filaments are of about
1 length unit in thickness.

The $z$-scrambled distribution then contains many points 
\begin{eqnarray}
g^j & \approx & (0,0)
\end{eqnarray}
which are strongly correlated [around the point $(0,0)$] on a scale
up to about 1 unit, even though the original distribution only contained
about one point in this region.

This was not a serious problem for the \citet{IovCS96} sample,
since all the survey dimensions were well above 
$\sim 100$--$200${\hMpc}.

Here, however, the 2QZ-10K patches 
(see fig.~1 of \citealt{Croom01}) are about
2 degrees in size, corresponding to $\sim 50$--$100${\hMpc} in 
the $0.6 < z < 2.2$ redshift range, depending on the values of 
the local cosmological parameters. It is likely that at least
some genuine filaments happen to lie in the radial and/or 
tangential directions, and some of the redshift selection effects
\citep{Scott91} are likely to create statistical excesses 
of quasars at certain favoured redshifts, i.e. creating
what could be called ``selection effect filaments'' 
in the tangential direction.

What is the thickness of a filament, as traced by quasars?
A minimum thickness is that 
corresponding to a few times 
the scale of the non-linear correlation length,
$r_0\sim5${\hMpc}, a maximum thickness is about 50\% of typical
void sizes, i.e. 50\% of $\sim 50$--$100${\hMpc},
i.e. $r \ltapprox 25$--$50${\hMpc}.	

This is the scale up to which an underestimate of $\xi(r)$ is
possible. On larger scales, there should be a smooth transition
to a slight overestimate of $\xi(r)$, due to pair conservation.

Since small scale correlations are subject to redshift evolution
and the effects of non-linear growth of structure, they would require
model-dependent interpretations if they were to be used 
for comoving standard ruler tests. So, the possible anti-correlation 
on small scales, 
due to $z$-scrambling in a survey with small dimensions, only affects
the part of $\xi(r)$ which is most difficult to use, and should not
be a problem for the present paper.



\section{Results} \label{s-results}

\fABC
\fABCb
\fABCc


\fABCwq



\subsection{Correlation functions}

Figs~\ref{f-corrnlowz}--\ref{f-corrnhizEdS} 
show the correlation functions $\xi(r)$ 
for the low-$z$, med-$z$ and hi-$z$ redshift intervals of the 2QZ-10K,
for the local metric parameter values 
($\Omm=0.3, \Omega_\Lambda=0.7)$ favoured from 
other observations, and for 
the now disfavoured Einstein-de Sitter 
metric ($\Omm=1.0, \Omega_\Lambda=0.0)$. 

Keeping in mind that the method of calculation of $\xi(r)$ 
adopted here,
which uses $z$-scrambling, 
is conservative and is likely to imply cancellation of some
real correlations, i.e. to underestimate the amplitudes of the
correlations, and that either real or selection effect structures
in the data are likely to cause an anti-correlation at small-scales,
as indicated by the shading at $r\ltapprox 50${\hMpc} in 
Figs~\ref{f-corrnlowz}--\ref{f-corrnhizEdS}, the following 
are clear from the figures:
\begin{list}{(\roman{enumi})}{\usecounter{enumi}}
\item 
A local maximum at a fixed comoving scale, 
$r\approx 240${\hMpc}, is clearly 
present in all three redshift ranges for
($\Omm=0.3, \Omega_\Lambda=0.7)$, at signal-to-noise ratios 
[defined in Eq.~(\ref{e-defnsn})]
of $(S/N)=2.4$, $(S/N)=1.1$ 
and $(S/N)=1.6$ 
at low-$z$,  med-$z$ and hi-$z$ respectively.
\item 
 A local maximum at a fixed comoving scale, 
$r\approx 180${\hMpc}, is 
present in the two higher redshift ranges for
($\Omm=0.3, \Omega_\Lambda=0.7)$, at a signal-to-noise ratio of
$(S/N)= 1.4$ at med-$z$ and $(S/N)= 1.7$ at hi-$z$, 
but absent in the low-$z$ data.
\item
A local maximum at a fixed comoving scale, 
$r\approx 90${\hMpc}, is 
present in the two higher redshift ranges for
($\Omm=0.3, \Omega_\Lambda=0.7)$, but absent in the low-$z$ data.
\item
The anti-correlation at $r\ltapprox 50${\hMpc} 
expected, due to ``aligned filaments'', the survey geometry and
$z$-scrambling,
is present at these scales,  for both choices of local cosmological
parameters. This implies a risk of systematic error
if the $r\approx 90${\hMpc} maximum were to be used for the 
consistency of comoving scale test, since it could be strongly
affected by this effect.
So, the conservative approach
of using $\rmin$ values of $100, 150$ and $200${\hMpc} will 
be retained here.
\item 
The low-$z$ correlation function
for ($\Omm=1.0, \Omega_\Lambda=0.0)$, 
Fig.~\ref{f-corrnlowzEdS}, clearly shows local maxima resembling
those in the low-$z$ 
($\Omm=0.3, \Omega_\Lambda=0.7)$ correlation function, but displaced to
smaller separations. In contrast, the med-$z$ and hi-$z$ 
correlation functions 
for ($\Omm=1.0, \Omega_\Lambda=0.0)$ only show broad, more or less
flat features.
\item 
The best sign of a feature at a consistent comoving scale for 
($\Omm=1.0, \Omega_\Lambda=0.0)$
is the $(S/N)\approx 2$ maximum at $r\approx
160${\hMpc} in the low-$z$ correlation function, and the $(S/N)\approx
0.5$ maximum at $r\approx 150${\hMpc} in the med-$z$ correlation
function.  Since low $(S/N)$ has not been used as a criterion for
excluding maxima, these two features could have, in principle, provided
an argument in favour of this choice of metric parameters, according
to the method defined above, if the feature were also present in the
hi-$z$ sample. However, it does not appear to be present in the hi-$z$
sample, and as is seen below quantitatively, is rejected as a maximum
present in all three redshift ranges.
\end{list}

\subsection{Comoving consistency of the position of a local maximum 
at all redshifts}
\label{s-consistency}

Figs~\ref{f-ABC}, \ref{f-ABCb} and \ref{f-ABCc} show the confidence
levels for rejecting the hypothesis that a local maximum is present
in all three redshift intervals, for $\rmin=100${\hMpc}, 
$\rmin=150${\hMpc} and 
$\rmin=200${\hMpc}, i.e. without pre-selecting a particular scale. 

For $\rmin=100${\hMpc} and $\rmin=150${\hMpc}, only one or two 
solutions exist. The solution in common to both, and also consistent
with the test for $\rmin=200${\hMpc}, is that with
($\Omm\approx0.35, \Omega_\Lambda\approx 0.55$). 
The $\rmin=200${\hMpc} test provides a much larger range of consistent
solutions, consistent with the flat, 
($\Omm=0.3, \Omega_\Lambda= 0.7$) metric, but favouring a 
hyperbolic metric [commonly called an 
``open'' metric, either ``open open'' or ``closed open'';
see \citet{LR99} or \citet{Rouk00c} 
to see why an ``open'' universe may be either 
open or closed].

\fcorrnlowzL

The reason for the different sizes of the 68\% contours in these
plots are clear from the correlation functions 
(Figs~\ref{f-corrnlowz}--\ref{f-corrnhizEdS}). The maximum at
$r\approx 240${\hMpc} for 
$(\Omm=0.3,\Omega_\Lambda=0.7)$
 is clearly present in all three redshift
ranges (Figs~\ref{f-corrnlowz}--\ref{f-corrnhiz}),
but in the low redshift range for these metric parameters 
(Fig.~\ref{f-corrnlowz}),
only slight, noisy ripples, much smaller than the error bars, are present above
100{\hMpc} and 150{\hMpc} for matching the $r\approx 180${\hMpc} maximum
in the med-$z$ and hi-$z$ samples.

Fig.~\ref{f-corrnlowzL} shows why a consistent solution is found
at ($\Omm\approx0.35, \Omega_\Lambda\approx 0.55$): a local maximum 
in the range $160 < r < 180${\hMpc} is present, though it is 
not significant
given the error bars. Both the $\rmin=100$ and $\rmin=150$ tests 
consider this peak to be consistent with the $r\approx 180${\hMpc}
in the med-$z$ and hi-$z$ redshift samples. It is clear that this
is not a significant solution.

It is also clear that no local maximum is common to all three redshift
ranges for the Einstein-de Sitter metric 
($\Omm=1.0, \Omega_\Lambda=0.0)$, nor for a low matter density model
with zero cosmological constant 
($\Omm=0.3, \Omega_\Lambda=0.0)$.

\subsection{What are the implied constraints on the metric parameters?}

Given that the $r\sim 180${\hMpc}
maximum providing the solution in Figs~\ref{f-ABC} and \ref{f-ABCb}
has a low signal-to-noise ratio, and that the $r\approx 240${\hMpc} 
clearly has the strongest signal-to-noise ratio in all redshift
ranges, the most robust constraint on the metric parameters 
is that of Fig.~\ref{f-ABCc}. 

This result is
\begin{eqnarray}
\Omm &=& 0.25\pm 0.10 \;\;(68\%\; \mbox{\rm confidence}) \nonumber \\
\Omm &=& 0.25\pm 0.15 \;\;(95\%\; \mbox{\rm confidence}) \nonumber \\
\Omega_\Lambda &=& 0.65\pm 0.25 \;\;(68\%\; \mbox{\rm confidence}) \nonumber \\
\Omega_\Lambda &=& 0.60\pm 0.35 \;\;(95\%\; \mbox{\rm confidence}) .
\label{e-ommLestimate}
\end{eqnarray}

The length scale of the local maximum is also constrained.
The 68\% and 95\% confidence intervals imply nearly identical 
estimates and uncertainties in the length scale of the local maximum,
which can be labelled $2\llss$ because of its similarity to twice the length
scale claimed in many other studies at low redshift. The scale is estimated
by taking the mean and standard deviation of the values $\rmax$ of the local
maximum for the three redshift intervals, either with or without weighting
by the rejection probability, within either the $1-P=68\%$ or the 
$1-P=95\% $ confidence contour. 

This formally yields
\begin{eqnarray}
2\llss &=& 242.9\pm 16.1{\hMpc} \;\;(68\%\; \mbox{\rm confidence}) \nonumber \\
2\llss &=& 243.0\pm 16.6{\hMpc} \;\;(95\%\; \mbox{\rm confidence}) \nonumber \\
2\llss &=& 244.4\pm 16.5{\hMpc} 
\;\;(68\%\; \mbox{\rm confidence, weighted}) \nonumber \\
2\llss &=& 244.3\pm 16.7{\hMpc} \;\;(95\%\; \mbox{\rm confidence, weighted})
\nonumber \\
& & 
\label{e-2llssestimate}
\end{eqnarray}
which can be summarised as
\begin{eqnarray}
\llss &=& 122\pm 9{\hMpc} \;\;(95\%).
\label{e-llssestimate}
\end{eqnarray}

 A similar hypothesis test to the above for $\rmin=200${\hMpc}, 
i.e. using 
the $r\approx 240${\hMpc} local maximum, but for a flat metric 
in an effective quintessence model has also been performed 
(Fig.~\ref{f-ABCwq}). 
This yields
\begin{eqnarray}
\Omm &=& 0.25\pm 0.10 \;\;(68\%\; \mbox{\rm confidence}) \nonumber \\
\Omm &=& 0.2\pm 0.2 \;\;(95\%\; \mbox{\rm confidence}) \nonumber \\
\wQ &<& -0.5 \;\;(68\%\; \mbox{\rm confidence})  \\
\wQ &<& -0.35 \;\;(95\%\; \mbox{\rm confidence}).
\label{e-wQestimate}
\end{eqnarray}
The introduction of external constraints would strengthen
these results, e.g. $\Omm \ge 0.2$ would imply 
$\wQ < -0.5 \;\;(95\%\; \mbox{\rm confidence})$, but for this study
it is preferred to deduce constraints independently of other 
observational analyses.

\fD

\fDp

\fDphun

\section{Discussion} \label{s-discuss}

Is the local maximum found at $2\llss=244\pm17${\hMpc} in all three
redshift ranges for $(\Omm\approx0.25, \Omega_\Lambda\approx 0.65)$
a real, cosmological signal or could it just be noise which happens
to give the signal expected? Answers to this question can be 
divided according to whether or not and which 
external information is accepted as a valid prior assumption.

There are strong observational justifications for expecting
that $\Omm\approx 0.3$, $\Omega_\Lambda\approx 0.7$, 
and there are also numerous observational analyses in favour
of fine features in the power spectrum or correlation of density
perturbations as traced by extragalactic objects, 
in particular, in favour of 
a local maximum at $\llss\approx 130\pm10${\hMpc}. 
The former are
close to being widely accepted by many independent groups, but 
the latter remain controversial. 

So, the reality of the signal can be discussed in 
the context of
\begin{list}{(\roman{enumi})}{\usecounter{enumi}}
\item no assumption regarding $\Omm, \Omega_\Lambda, \llss$; or
\item the assumption that $\Omm\approx 0.3, \Omega_\Lambda\approx0.7$, 
but no assumption regarding $\llss$; or
\item the assumption that $\llss\approx 130\pm10${\hMpc},
but no assumptions regarding $\Omm, \Omega_\Lambda$; or
\item the assumptions that 
$\Omm=0.25, \Omega_\Lambda= 0.65$ and
$2\llss= 244\pm17${\hMpc}.

\end{list}

\subsection{(i) No assumption regarding $\Omm, \Omega_\Lambda, \llss$}

Let us make no assumption regarding $\Omm, \Omega_\Lambda$ and $\llss$.

Consider the dependence of an arbitrary property of $\xi(r)$, 
represented by some statistic of $\xi(r)$,
as a function of $\Omm$ and $\Omega_\Lambda$. 

The value of $\xi(r)$ in any bin depends on the numbers of 
pairs falling in the bin, i.e. on the numbers of 
``data-data'' (DD)  pairs, ``data-random'' (DR) 
pairs, and ``random-random'' (RR) pairs. 

For a fixed distribution 
in redshift and angular position, as $\Omm$ decreases and/or
as $\Omega_\Lambda$ increases, the proper distance separation 
between any given pair of quasars increases. So, 
as $\Omm$ decreases and/or as $\Omega_\Lambda$ increases,  
the full distribution of pair separations is stretched over
a larger interval of proper separations,
and the Poisson error per bin increases. This stretching is mostly
in the radial direction, since the radial sizes of the densely observed
regions of the 2QZ-10K are about an order of magnitude greater than
their tangential sizes.

This can be shown algebraically as follows. Ignoring the angular 
separations, since these are small, the full pair distribution is
spread over the interval which is approximately
\begin{equation}
0 \le r \le d(\Omm,\Omega_\Lambda,z_2)-d(\Omm,\Omega_\Lambda,z_1)
\label{e-fulldist}
\end{equation}
where 
$d(\Omm,\Omega_\Lambda,z_i)$ are the proper distances from the observer
to the redshift limits $z_i$.
The number of pairs $m$ in any given
separation bin is
\begin{eqnarray}
m & \propto & { 1 \over 
d(\Omm,\Omega_\Lambda,z_2)- 
d(\Omm,\Omega_\Lambda,z_1) }.
\label{e-onebin}
\end{eqnarray}
So, the fractional Poisson error per bin is
\begin{eqnarray}
{\Delta m \over m} & \propto & \sqrt{ 
d(\Omm,\Omega_\Lambda,z_2)- 
d(\Omm,\Omega_\Lambda,z_1) }.
\label{e-noiseonebin}
\end{eqnarray}
Thus,
the Poisson error per bin increases when 
$\Omm$ decreases and/or $\Omega_\Lambda$ increases
[see eqs~(\ref{e-defdprop}), (\ref{e-defcurv})].

If $\xi(r)$ has no fine scale features, i.e. only noise fluctuations,
then the increase in Poisson error
should be the only systematic effect of changing $\Omm$ and/or
$\Omega_\Lambda$, and it should be a {\em monotonic} effect, in 
the sense that the highest amount of noise should occur 
for the lowest $\Omm$ and the highest $\Omega_\Lambda$ calculations.

The favoured values of $\Omm$ and $\Omega_\Lambda$ found above
[Eq.~(\ref{e-ommLestimate})]
are {\em not} the lowest and highest (respectively) of the domain
investigated. This provides one argument against the 
local maximum at $2\llss=244\pm17${\hMpc} being a noise fluctuation.

An independent argument is to define a difference statistic
\begin{eqnarray}
D_0(\Omm,\Omega_\Lambda) &\equiv& 
\sum_{r_1}^{r_2}
|\xi(r+\Delta r,\Omm,\Omega_\Lambda)-\xi(r,\Omm,\Omega_\Lambda)| 
\nonumber \\ 
& & 
\label{e-D0}
\end{eqnarray}
and its normalised value
\begin{eqnarray}
D(\Omm,\Omega_\Lambda) &\equiv& 
{D_0(\Omm,\Omega_\Lambda) \over
\max_{\Omm,\Omega_\Lambda} \{ D_0(\Omm,\Omega_\Lambda) \} },
\label{e-D}
\end{eqnarray}
over a range of pair separations $r_1 \le r \le r_2.$
The limits $r_1, r_2$ can be chosen in order to answer the question 
raised above regarding the $2\llss$ local maximum, i.e.
$r_1=200${\hMpc},
$r_2=300${\hMpc}, or can be chosen to further remove prior
knowledge of $2\llss$ while remaining in a domain where noise
signals are least likely to occur, i.e. 
$r_1=100${\hMpc}, $r_2=350${\hMpc}.

This statistic does {\em not} contain any information on whether or not
a local maximum is present in this range. It can be said 
to represent the average absolute
value of the slope of $\xi(r)$, or more informally, how much $\xi(r)$
fluctuates.

If only noise is present, then $D$ should vary monotonically as a
function of $\Omm$ and $\Omega_\Lambda$, since the slopes of $\xi(r)$
would only be noise-generated.

On the other hand,
if a cosmological fluctuation in $\xi(r)$ is present in the data, 
then whether or not it's a local 
maximum, a local minimum or just the rising or falling slope of
one or the other, the values of $D$ near the correct values of
$(\Omm,\Omega_\Lambda)$ should be higher than for surrounding
values of $(\Omm,\Omega_\Lambda)$.

An approximate correction for the monotonic variation in $D$,
due to the decreased numbers of pairs per bin
and resulting increased (fractional) Poisson errors for lower
$\Omm$ and higher $\Omega_\Lambda$ values, can be made 
by using dependence of the maximum proper distance,
$d(\Omm,\Omega_\Lambda,z)$, across
the full redshift range, and
replacing $D$ by 
\begin{eqnarray}
&&D'(\Omm,\Omega_\Lambda) \equiv \nonumber \\
&&\;\;{D(\Omm,\Omega_\Lambda) 
 { \left[
{
d(\Omm,\Omega_\Lambda,2.2) - d(\Omm,\Omega_\Lambda,0.6) 
\over 
d(0.0,1.1,2.2) - d(0.0,1.1,0.6)} 
 \right]^{-1/2} }
}.  \nonumber \\
&&
\label{e-Dp}
\end{eqnarray}

Fig.~\ref{f-D} shows the dependence of $D$ on $\Omm$ and $\Omega_\Lambda$.

Instead of $D$ increasing smoothly and monotonically from 
$(\Omm=1.1,\Omega_\Lambda=-0.1)$ to $(\Omm=0.0,\Omega_\Lambda=1.0)$,
as would be the case if fluctuations were only caused by noise,
it is clear that $D$ does {\em not} vary monotonically with
$\Omm$ and  $\Omega_\Lambda$. 

Moreover, the exceptional region is clearly 
that of the excess values of $D$ 
surrounding $(\Omm\approx 0.25, \Omega_\Lambda\approx0.55$). It is not possible
that the region $\Omm\approx 0.15, \Omega_\Lambda\approx 0.8$ is a dip
relative to the noise values of $D$, since any additional component to
$D$ not caused by noise must cause a positive contribution (absolute
values are always non-negative).

The approximate correction for the monotonic dependence on Poisson noise
[Eq.~(\ref{e-Dp})]
shows this excess signal even more clearly, in $D'$, shown in
Figs~\ref{f-Dp} and \ref{f-Dp100}. 
Although the Poisson noise dependence is not completely
removed, since the correction applied is only approximate, it is clear
that $D'$ has excessively large values in the linear region 
from $(\Omm\approx 0.25, \Omega_\Lambda\approx0.55$) to
$(\Omm\approx 0.1, \Omega_\Lambda\approx0.3$), and that this is the
case whether or not a relatively small region around the 
$2\llss=244\pm17${\hMpc} maximum, i.e. 
$200${\hMpc}~$ \le r \le 300${\hMpc} 
(Fig.~\ref{f-Dp}), or 
a relatively large region around the maximum,
i.e. $100${\hMpc}~$ \le r \le 350${\hMpc} 
(Fig.~\ref{f-Dp100}), is used.

It is difficult to explain these excess values of $D'$ unless
{\em there is a cosmological signal for some $(\Omm,\Omega_\Lambda)$ pair
in the degeneracy region 
from $(\Omm\approx 0.25, \Omega_\Lambda\approx0.55)$ to
$(\Omm\approx 0.1, \Omega_\Lambda\approx0.3)$.}

Suppose a metric value pair outside of this region, 
e.g. $(\Omm\approx 1.0, \Omega_\Lambda\approx0.0)$, is valid. 
Then, either 
\begin{list}{(\alph{enumi})}{\usecounter{enumi}}
\item there is a fluctuation of cosmological origin 
in $\xi(r)$ of the 2QZ-10K, or 
\item there is not, i.e. there is only noise.
\end{list}

Consider each case.
\begin{list}{(\alph{enumi})}{\usecounter{enumi}}
\item If $(\Omm\approx 1.0, \Omega_\Lambda\approx0.0)$
and there is a cosmological fluctuation in $\xi(r)$, then there must 
be an excess value of $D$ and $D'$ near 
$(\Omm\approx 1.0, \Omega_\Lambda\approx0.0$), since for any 
{\em wrong} value of
$(\Omm,\Omega_\Lambda)$, the ``correlated pairs'' which would
have been placed in a single separation
bin, say, the $i^{\mbox{\small th}}$ separation bin,  
$r_1 + (i-1)\Delta r < r < r_1 +i \Delta r$ 
[for $(\Omm\approx 1.0, \Omega_\Lambda\approx0.0)$],
are diluted over several bins 
$\{ r_1 + (i'-1)\Delta r < r < r_1 +i' \Delta r 
\}_{i'=i,i-1,i-2,\ldots,i+1,i+2,\ldots}$ 
and the value of 
$D$ and $D'$ is decreased. However, there is {\em not} an 
excess in $D$ and $D'$ at $(\Omm\approx 1.0, \Omega_\Lambda\approx0.0)$.
So, case (a) is wrong.
\item 
If $(\Omm\approx 1.0, \Omega_\Lambda\approx0.0)$ and 
and there is {\em no} cosmological fluctuation in $\xi(r)$, 
then the fluctuations in $\xi(r)$ are Poisson noise fluctuations.
This noise must increase monotonically as $\Omm$ decreases and/or as
$\Omega_\Lambda$ increases [Eq.~(\ref{e-noiseonebin})].
Hence, there must {\em not} be an excess of $D$ and $D'$ near
$(\Omm\approx 0.25, \Omega_\Lambda\approx0.55)$, since this is far
from the limit $(\Omm=0.0, \Omega_\Lambda=1.0)$. 
However, there {\em is}  an excess of $D$ and $D'$ near
$(\Omm\approx 0.25, \Omega_\Lambda\approx0.55)$. So,
case (b) is wrong.
\end{list}

Since both (a) and (b) are wrong, 
$(\Omm\approx 1.0, \Omega_\Lambda\approx0.0)$ 
is invalid independently of whether or not the existence of 
a fluctuation of cosmological origin is assumed.

The same argument holds for other pairs 
$(\Omm, \Omega_\Lambda)$ outside of the degeneracy region 
from $(\Omm\approx 0.25, \Omega_\Lambda\approx0.55)$ to
$(\Omm\approx 0.1, \Omega_\Lambda\approx0.3)$.

The only reasonable 
interpretation of Figs~\ref{f-D}--\ref{f-Dp100} is that 
$(\Omm,\Omega_\Lambda)$ lies 
in the degeneracy region 
from $(\Omm\approx 0.25, \Omega_\Lambda\approx0.55)$ to
$(\Omm\approx 0.1, \Omega_\Lambda\approx0.3)$ and that $\xi(r)$ 
has at least some fluctuating component which is a comoving, cosmological
signal and not Poisson noise.

\subsection{(ii) The assumption 
that $\Omm\approx 0.3, \Omega_\Lambda\approx0.7$, 
but no assumption regarding $\llss$}

Given the prior estimate that $\Omm\approx 0.3, \Omega_\Lambda\approx0.7$, 
Figs.~\ref{f-corrnlowz}, \ref{f-corrnmedz} 
and \ref{f-corrnhiz} show that a local maximum of signal-to-noise
ratio [Eq.~(\ref{e-defnsn})] of $S/N\approx 3$, 1 and 1 exists 
near $2\llss=244\pm17${\hMpc} in the low, medium and high redshift ranges
respectively.

Why should a local maximum due to noise occur at the same position in
all three redshift ranges, to within the precision of one bin 
($\Delta r =5${\hMpc}), if it were not physical? 

The interval between successive maxima is at least $\approx 60${\hMpc}.
In the low redshift bin the interval is higher, since the larger error bars
lead to insignificant maxima at lower length scales, but let us be
conservative and adopt $60${\hMpc} as the scale along which a random
maximum could occur. If it is assumed that local maxima occur about once
within this interval but with random phase, then
the chance of agreement between two redshift 
ranges to within $\pm 5${\hMpc} 
is then $P=10/60$, and between all three is 
$P=(10/60)^2 \approx 0.03$, i.e. the hypothesis of uniform 
random phases for these local maxima is rejected at the 
97\% confidence level.

The assumption  that $\Omm\approx 0.3, \Omega_\Lambda\approx0.7$ 
also suggests that the local maxima which appear at $r \approx 100${\hMpc} 
and $r\approx 180${\hMpc} in the
med-$z$ and hi-$z$ samples could be real. 

The former is likely to be affected by the existence of correlated
structures in a survey geometry of narrow dimensions and the use of
$z$-scrambling (\SS\ref{s-zscramprob}), so is unlikely to be
cosmological in origin. The latter should be considered a better
candidate for a real local maximum. However, given the lack of
its detection in the low-$z$ sample, it is more prudent to 
consider the 2QZ-10K data to be
insufficient to establish its existence.

\subsection{(iii) The assumption that $\llss\approx 130\pm10${\hMpc},
but no assumptions regarding $\Omm, \Omega_\Lambda$ }

If the existence of a local maximum near  $\llss\approx 130\pm10${\hMpc}
is assumed,
then in the analysis discussed 
in \SS\ref{s-consistency}
and shown in Figs~\ref{f-ABC} and \ref{f-corrnlowzL}, a search for a
consistent local maximum in 
the $(\Omm,\Omega_\Lambda)$ plane for $\rmin=100${\hMpc} should have
led to a significant solution. 
However, 
only a weak solution exists, and Fig.~\ref{f-corrnlowzL} shows that this
is for a local maximum at $160 < r < 180${\hMpc} rather than at 
$130\pm10${\hMpc}, with a negligible signal-to-noise ratio.

At first sight, this is, therefore, the most puzzling result of this
analysis. If there are genuine fine features significantly detected in
$\xi(r)$ of the 2QZ-10K, why should the most commonly claimed feature
at low redshift be absent?

Two likely answers include:
\begin{list}{(\roman{enumi})}{\usecounter{enumi}}
\item the sparsity of the sample, and/or
\item coincidence of genuine correlations with redshift and or angular selection
effects 
\end{list}
which are both capable of removing underlying matter density correlations
from the sample. No amount of correction can (validly) re-introduce 
correlations which are absent from a catalogue due to one or both of
these reasons.

Another possibility is:
\begin{list}{(\roman{enumi})}{\usecounter{enumi}}
\addtocounter{enumi}{2}
\item that by use of $z$-scrambling and observational angular 
positions to mimic
selection effects as closely as possible, the cosmological 
correlations have also been
mimicked in the random catalogues and, hence, been cancelled out of $\xi(r)$.
\end{list}

Note that these effects are related to the use of a conservative
technique: careful correction for selection effects risks
cancelling some real, cosmological correlations, 
but implies that any detected signal is very 
unlikely to be a selection effect.

If one or several of the explanations (i)--(iii) are correct, then it
should still be expected that if a local maximum is detected at a
scale where sparsity or selection effects are less serious, it is
detected at a harmonic of the assumed scale.

This is indeed the case.
Without prior constraints on $(\Omm, \Omega_\Lambda)$, 
a local maximum at the scale $2\llss=244\pm17${\hMpc} was detected.

This supports the claim that a local maximum in $\xi(r)$ or the
power spectrum exists at $\llss\approx 130\pm10${\hMpc}, and 
possibly provides a more accurate estimate 
($\llss=122\pm9${\hMpc}) than the low
redshift estimates, since
the effects of peculiar velocities are much smaller than for the
low redshift samples.

Earlier arguments related to (i) include the suggestion 
by \citet*{deLapp91} 
that the existence of a scale of $\llss\sim120${\hMpc}, traced by
positive density fluctuations, is implied
by the observed existence of large scale structure 
at very low redshifts 
\citepf{deLapp86}, consisting of
negative density fluctuations on a scale of around 20--50{\hMpc}.

\subsection{(iv) The assumptions that 
$\Omm= 0.25, \Omega_\Lambda= 0.65$ and
$2\llss= 244\pm17${\hMpc}}

Even if the values of the metric parameters are assumed and a 
scale is chosen, there is no reason why a Poisson noise peak should
occur at this particular 
scale (claimed to be of interest in several observational
analyses) in all three redshift intervals, 
in excess of Poisson noise peaks generated from uncorrelated,
random simulations, unless it is cosmological in origin, provided
that the simulations mimick any possible peculiar properties of
the catalogue.

So, 200 random simulations, using the $z$-scrambling 
technique as above, were performed for each redshift interval, for
each angular sub-sample, for 
($\Omm= 0.25, \Omega_\Lambda= 0.65$).

Similarly to \citet{RM01}, 
the probability $P_i$ that a local maximum can occur 
as close to 
$2\llss= 244\pm17${\hMpc} in the simulation as in the observational
catalogue, 
with at least the same signal-to-noise 
ratio [Eq.~(\ref{e-defnsn})] as the observational local maximum, 
for the $i^{\mbox{\small th}}$ redshift interval,  is defined 
\begin{eqnarray} 
P_i &\equiv &
P{\left[ 
{\rule[-0.3ex]{0ex}{3ex}\;} 
|\rsim
-2\llss| \le|\robs-2\llss|^{\mbox{\rm \ }}  
 \right.}  \anddd \nonumber \\
&& \left. (S/N)^{\mbox{\rm sim}} \ge (S/N)^{\mbox{\rm obs}} 
{\;\rule[-0.3ex]{0ex}{3ex}} 
\right], 
\label{e-defnpsim} 
\end{eqnarray}
where the position and signal-to-noise ratio of the first 
local maximum at $r > \rmin=200${\hMpc} are 
$\robs$ and $(S/N)^{\mbox{\rm obs}}$ in the observations 
and $\rsim$ and $(S/N)^{\mbox{\rm sim}}$ in the simulations.
The combined probability for the three redshift
intervals is
\begin{eqnarray} 
P_{123} &= & P_1 P_2 P_3
\label{e-defnpsimall} 
\end{eqnarray}
since the observational data sets in 
the three redshift intervals are independent. 

The results are $P_1=P_2 = 0.065, P_3=0.075$, so that
$P_{123}=3\e{-4}$. 

In other words, given that 
$(\Omm= 0.25, \Omega_\Lambda= 0.65)$,
the probability that random Poisson signals 
give a local maximum as close to 
$2\llss= 244${\hMpc} as the local maximum in the observational data and of
at least as strong a signal-to-noise ratio as the observational 
local maximum, in all three redshift
intervals, is rejected at the $99.97\%$ confidence level.

\subsection{Comparison with the analysis of \protect\citet{Hoyle01}}

How do these results compare with the fourier spectrum analysis
by \citet{Hoyle01}?

Given the large redshift range of the 2QZ-10K (80\% of the quasars lie 
in the range $0.6 \ltapprox z\ltapprox 2.2$), it is puzzling that
\citet{Hoyle01} found a local maximum to exist for widely differing choices
in the local cosmological parameters: 
it appears to be present both for $(\Omm=0.3, \Omega_\Lambda=0.7)$ and 
for $(\Omm=1.0, \Omega_\Lambda=0.0)$, at 
$r\approx 90${\hMpc} and $r\approx 65${\hMpc} respectively in their
analysis.

Comparison of fig.~8(a) and fig.~8(b) of \citet{Hoyle01},
representing the cases
$(\Omm=1.0, \Omega_\Lambda=0.0)$ and
$(\Omm=0.3, \Omega_\Lambda=0.7)$ respectively, seems to show a stronger
signal in the latter case, but this is still problematic for a feature
which should exist at a fixed comoving scale only for the correct
values of the local cosmological parameters.

However, differences in the technique of \citet{Hoyle01} relative to our
own include:
\begin{list}{(\roman{enumi})}{\usecounter{enumi}}
\item \citeauthor{Hoyle01}'s use of a very smooth redshift selection function,
which probably does not correct for the selection effects discussed
by \citet{Scott91}; 
\item \citeauthor{Hoyle01}'s inclusion of angular selection regions 
that have
less than 80\% ``coverage completeness''
(two-thirds of their sample lies in regions
with only 20\%--80\% coverage completeness),
with a statistical
correction for this via their window function;
\item use of a fourier analysis rather than a correlation function;
\item use of large, logarithmic bins in $k$ rather than linear 
bins of $\Delta r=5${\hMpc} in $r$.
\end{list}

(i) The non-removal of redshift selection effects seems a reasonable 
candidate for inducing a convolution of an artefact with the real signal,
shifting it to the wrong scale, and generating a weak signal for the wrong
values of the metric parameters.

(ii) Inclusion of low ``coverage completeness'' regions is likely to 
introduce strong sources of noise.
Even though this should be statistically
corrected by the window function, our own results suggest that it may
be more useful simply to ignore low ``coverage completeness'' regions. 
The disadvantage is that only about 3000 of the 11000 quasars can be used,
but the advantage is having less noise to correct.

(iii) A fourier analysis of data with a very complex window function is
obviously difficult, and probably more useful for estimating the overall
shape of the power spectrum than for the detection of fine features in it.
Although in principle, the 
\dobold{power} spectrum and the correlation function are 
intrinsically related \dobold{through Fourier transforms}, 
conversion from one to another
is difficult in practice and will be left to other authors, particularly
since a large part of the power is removed from our estimate of $\xi(r)$
by the use of $z$-scrambling: this technique is optimised to detect
fine features, not gross features.

\dobold{It should be noted that in practice, the correlation function
is more commonly calculated on small scales, while the power spectrum
is calculated on large scales.  This is probably because the simplest
implementation of calculation of the correlation function, via direct
counting of pairs, is more computationally intensive than that of 
fast fourier transforms.

However, if ($\Omm,\Omega_\Lambda$) parameter space is to be explored,
then using correlation functions is more precise than using Fourier
transforms. This is because Fourier transforms don't exist in curved
space. In curved space over the range considered, calculation of
Fourier transforms instead of 3-spherical harmonics or 3-hyperbolic
eigenfunctions is still a good approximation up to a few hundred Mpc,
since the curvature radius is at least 2860{\hMpc}. Moreover, 
observations indicate that the curvature radius is at least
about 10,000{\hMpc}.}


(iv) The use of wide, logarithmic bins may make it difficult to detect
a feature which (in the present analysis) is much less strong 
in amplitude than that claimed by \citet{Einasto97nat}.

Any or all of these could explain the differences in the two analyses.

It should be noted that \citet{Croom00} performed a correlation
function analysis of the 2QZ-10K on scales below $\sim 100${\hMpc},
avoiding problem (iii) above, making calculations to consider the
effects of (ii), but retaining problems (i) and (iv). 
Problem (i) is likely to be insignificant on scales below
$\sim 100${\hMpc}, so should not be 
a problem for the purposes of the analysis 
of \citet{Croom00}.

In this paper, scales above 100{\hMpc} are those of most interest,
which is why the correction technique of $z$-scrambling has been used
here to avoid problem (i). So, the two
studies are complementary and cannot be directly compared.

\subsection{Comparison with the analyses of \protect\citet{RM00,RM01}}

As mentioned above (P.~Petitjean, private communication), 
some fraction of the \citet{IovCS96} sample of high-quality 
quasar candidates 
appears to consist of
quasars at wrongly estimated redshifts and stars.
This implies that some ``redshifts'' in the analyses either are those of
genuine quasars, but for a misidentifed emission line,
or correspond to the emission wavelengths of atomic transitions which 
should vary little from star to star. 

Both cases should add random noise in the cross-correlations, and
some genuine correlations should exist among the misidentified quasars, 
but at underestimated tangential separations (if the true redshifts
are lower than the estimated ones, which is apparently the case).

However, since the emission lines of the stars should be at essentially
fixed ``false redshifts'' (since the velocities of stars are much less than
cosmological expansion velocities), the strongest artefact in the 
sample is likely to be objects at certain favoured ``false redshifts''.

In this case, the purely tangential analysis \citep{RM00} is likely
to be less affected by contaminating objects 
than the three-dimensional analysis \citep{RM01}. 
Since the scale found in \citet{RM00} 
is consistent with the present value of 
$\llss\approx 122\pm9${\hMpc} for values of $\Omm$ consistent with those
estimated here, the results are compatible. 

\subsection{Possible theoretical explanations for local maxima in 
$\xi(r)$}

Possible theoretical explanations for fine features in the correlation
function or power spectrum include:
\begin{list}{(\roman{enumi})}{\usecounter{enumi}}
\item acoustic 
baryonic fluctuations formed during the transition of the
recombination epoch, e.g. \citet*{Eisen98a,MeikWP98,Peeb99b,MillerNB01};
\item baryonic fluctuations formed during 
an inflationary epoch, 
as a result of the evolution of a complex scalar field condensate, 
\citet{KirCh00}; 
\dobold{ \item 
phase transitions in supersymmetric models of double inflation,
(e.g. \citealt{Lesg98,Lesg00});
} or
\item string theory inspired fluctuations formed during the Planck epoch.
\citet{MB00b,Easth01}
\end{list}

It would clearly be premature to attempt to distinguish between these on the
basis of the 2QZ-10K data.

\section{Conclusions} \label{s-conclu}

Although considerable care is required to avoid selection effects
in the ``10K'' initial release of the 2dF QSO Redshift Survey (2QZ-10K), 
it is difficult to avoid the conclusion that a harmonic of the
$\llss\approx 130\pm10${\hMpc} local maximum in the correlation function
or power spectrum of density perturbations, as traced by extragalactic
objects, is present in the data and that this scale provides an
extremely model-free way of constraining local cosmological parameters 
without requiring combination with external data sets.

This local maximum  was found 
by estimating the spatial two-point autocorrelation functions $\xi(r)$
of the three-dimensional (comoving, spatial) distribution of
the $N=2378$ quasars in the most completely observed 
($\ge 85\%$ spectroscopic completeness) and ``covered'' 
($\ge 80\%$ coverage completeness)
sky regions of the catalogue, over the 
redshift ranges 
$0.6 < z < 1.1$ (``low-$z$''),
$1.1 < z < 1.6$ (``med-$z$''), and
$1.6 < z < 2.2$ (``hi-$z$''), using the $z$-scrambling technique
in order to avoid selection effects. 

\begin{list}{(\roman{enumi})}{\usecounter{enumi}}
\item Avoiding {\em a priori} estimates of the length scales of features, 
local maxima in $\xi(r)$ are found in the different redshift ranges.
The requirement that a local maximum be 
present in all three redshift ranges
at the same comoving length 
scales implies strong,
purely geometric constraints on the local cosmological parameters.
The only local maximum satisfying this requirement is that
at a length scale of $2\llss= 244\pm17${\hMpc}.
\item
 For a standard cosmological constant FLRW model, the matter density 
and cosmological constant are constrained to 
$\Omm= 0.25\pm0.10, \Omega_\Lambda=0.65{\pm0.25} $ (68\% confidence),
$\Omm= 0.25\pm0.15, \Omega_\Lambda=0.60{\pm0.35} $ (95\% confidence),
respectively, {\em from the 2QZ-10K alone}.
\item
 For an effective quintessence ($\wQ$) model and zero curvature, 
$ \wQ<-0.5 $ (68\% confidence),
$ \wQ<-0.35 $ (95\% confidence) are found,
again {\em from the 2QZ-10K alone}.
\end{list}

These results are consistent with 
type Ia supernovae and 
microwave background results
(e.g. \citealt{SCP9812,HzS98,Boom00a,Maxima00a}), but avoid the
need to combine two data sets in order to obtain constraints 
on both $\Omm$ and $\Omega_\Lambda$.

The constraints on $\Omm$ are, of course, 
in remarkable agreement
with the constraints from kinematics of galaxy
clusters (e.g. \citealt*{CYE97}), collapsing galaxy groups
\citep{Mam93}, as well as from the baryonic fraction in clusters
\citepf{WNEF93,Mohr99} and groups \citep{HMam94}.

However, in contrast to \citet{Jaffe00}, who exclude
a flat universe at about $\sim95\%$ significance, 
i.e. $\Omm+\Omega_\Lambda = 1.11^{+0.13}_{-0.12}$ at 95\% confidence,
a hyperbolic universe is weakly favoured here, though a flat universe 
is consistent with the data at the 68\% confidence level.

The present results also contrast with the constraint from 
the quietness of the Hubble flow (\citealt{Chemin00} and 
\SS5 of \citealt{Sand99}) which requires 
a cosmological constant of about $\Omega_\Lambda = 0.8$ or higher.

It should be emphasised that the 2QZ-10K data establish the existence
of a non-zero cosmological constant independently of the supernovae
results: $\Omega_\Lambda=0$ is refuted at the 99.7\% confidence level.

%

The full 2QZ will clearly provide even more impressive constraints 
on local maxima, $\Omm,$ $\Omega_\Lambda$ and $\wQ$.


%
%
%
%
%
%



\section*{Acknowledgments}

Useful comments from Patrick Petitjean, Brian Boyle and Scott Croom
are gratefully acknowledged. 
The 2QZ-10K was based on observations made with the Anglo-Australian 
Telescope and the UK Schmidt Telescope. 
Use of the
resources at the Centre de Donn\'ees astronomiques de Strasbourg 
({\em http://csdweb.u-strasbg.fr}),
the support of the Institut d'Astrophysique de Paris, CNRS, 
for a visit during which part of this work was carried out,
and the support of la Soci\'et\'e de Secours des Amis des Sciences
are gratefully acknowledged. This research has been supported by the 
Polish Council for Scientific Research Grant
KBN 2 P03D 017 19
and has benefited from 
the Programme jumelage 16 astronomie 
France/Pologne (CNRS/PAN) of the Minist\`ere de la recherche et
de la technologie (France).

\subm \clearpage ::::


\begin{thebibliography}{}
\bibitem[Adelberger et al.(1998)]{Adel98} \joref{Adelberger K.~L., Steidel C.~C., Giavalisco M., Dickinson M., Pettini M., Kellogg M.}{\apj}{508}{18}{1998}

\bibitem[Arnouts et al.(1999)]{Arno99} \joref{Arnouts S., Cristiani S., Moscardini L., Matarrese S., Lucchin F., Fontana A., Giallongo E.}{\mnras}{310}{540}{1999}

\bibitem[Balbi et {al.}(2000)]{Maxima00a} \joref{Balbi A. et al. (Maxima collaboration)}{{\apj}}{545}{L1}{2000} \ {(arXiv:astro-ph/0005124)}


\bibitem[Baugh \& Efstathiou(1993)]{BauE93} \joref{Baugh C.M., Efstathiou G.}{\mnras}{265}{145}{1993}
\bibitem[Baugh \& Efstathiou(1994)]{BauE94} \joref{Baugh C.M., Efstathiou G.}{\mnras}{267}{323}{1994}


\bibitem[Boyle et {al.}(2000)]{Boy00a} \joref{Boyle~B.~J., Shanks~T., Croom~S.~M., Smith~R.~J., Miller~L., Loaring~N., Heymans~C.}{\mnras}{in press}{}{2000} \ (arXiv:astro-ph/0005368)

\bibitem[Broadhurst(1999)]{Bro99} Broadhurst T., 1999, in `Clustering at High Redshift', ed. V.~Le~Brun, A.~Mazure, O.~Le~F\`evre, in press 

\bibitem[Broadhurst et al.(1990)]{Bro90} \joref{Broadhurst T. J., Ellis R. S., Koo D. C., Szalay A. S.}{Nature}{343}{726}{1990}
\bibitem[Broadhurst \& Jaffe(1999)]{BJ99} {Broadhurst T., Jaffe A.~H.}, {1999}, submitted \ (arXiv:astro-ph/9904348)

\bibitem[Carlberg et {al.}(1997)Carlberg, Yee \& Ellingson]{CYE97} \joref{Carlberg R.~G., Yee H.~K.~C., Ellingson E.}{\apj}{478}{462}{1997}

\bibitem[Chemin(2000)]{Chemin00} Chemin~A., 2000, Adv.~Sp.~Res., submitted



\bibitem[Croom et {al.}(2000)]{Croom00} \joref{Croom,~S.~M., Shanks,~T., Boyle,~B.~J., Smith,~R.~J., Miller,~L., Loaring,~N.~S., Hoyle,~F.}{\mnras}{}{}{2000} \ (arXiv:astro-ph/0012375) 

\bibitem[Croom et {al.}(2001)]{Croom01} \joref{Croom,~S.~M., Smith,~R.~J., Boyle,~B.~J., Shanks,~T., Loaring,~N.~S., Miller,~L., Lewis,~I.~J.}{\mnras}{322}{L29}{2001} \ (arXiv:astro-ph/0104095) 

\bibitem[da Costa(1992)]{daCosta92} da~Costa L.~N., 1992, in The Distribution of Matter in the Universe, ed. G.~A.~Mamon, D.~Gerbal (Meudon: Obs. de Paris), p163, ftp://ftp.iap.fr/ pub/from\_users/gam/PAPERS/DAECMTG/ \\ dacosta.dvi.Z

\bibitem[da Costa et {al.}(1993)]{daCosta93} da~Costa L.~N. et {al.}, 1993, in Cosmic Velocity Fields, ed. Bouchet, F., Lachi\`eze-Rey, M., (Gif-sur-Yvette, France: Editions Fronti\`eres), p475




\bibitem[de Lapparent et {al.}(1986)]{deLapp86} \joref{de~Lapparent V., Geller M.~J., Huchra J.~P.}{\apj}{302}{L1}{1986}

\bibitem[de Lapparent et {al.}(1991)]{deLapp91} \joref{de~Lapparent V., Geller M.~J., Huchra J.~P.}{\apj}{369}{273}{1991}

\bibitem[Deng et {al.}(1994)Deng, Xiaoyang \& Fang]{Deng94} \joref{Deng Z., Xiaoyang X., Fang L.-Zh.}{\apj}{431}{506}{1994} 
\bibitem[Deng et {al.}(1996)Deng, Deng \& Xia]{Deng96} \joref{Deng X.-F., Deng Z.-G., Xia X.-Y.}{Chin.Astron.Astroph.}{20}{383}{1996}


\bibitem[Easther et {al.}(2001)]{Easth01} \epref{Easther,~R., Greene,~B.~R., Kinney,~W.~H., Shiu,~G.}{arXiv:hep-th/0104102}{2001} 

\bibitem[Efstathiou(1999)]{Efst99} \joref{Efstathiou, G.}{\mnras}{310}{842}{1999} \ (arXiv:astro-ph/9904356)

\bibitem[Einasto et {al.}(1994)]{Einasto94} \joref{Einasto M., Einasto~J., Tago E., Dalton G. B., Andernach H.}{\mnras}{269}{301}{1994}

\bibitem[Einasto et {al.}(1997a)]{Einasto97corr} \joref{Einasto, J., et al.}{\mnras}{289}{801}{1997}

\bibitem[Einasto et {al.}(1997b)]{Einasto97nat} \joref{Einasto, J., et al.}{Nature}{385}{139}{1997}

\bibitem[Eisenstein(1998)]{Eisen98a} \joref{Eisenstein~D.~J., Hu~W., Silk~J., Szalay~A.~S.}{\apj}{494}{L1}{1998} 





\bibitem[Gazta\~naga \& Baugh(1998)]{GazB98} \joref{Gazta\~naga E., Baugh C.M.}{\mnras}{294}{229}{1998}

\bibitem[Geller \& Huchra(1989)]{GH89} \joref{Geller M.~J., Huchra J.~P.}{Science}{246}{897}{1989} 

\bibitem[Groth \& Peebles(1977)]{GroP77} \joref{Groth E.~J., Peebles P.~J.~E.}{\apj}{217}{385}{1977}


\bibitem[Guzzo(1999)]{Guzzo99} Guzzo L., 1999, in proceedings of XIX Texas Symp. Rel. Astr. \ (arXiv:astro-ph/9911115)

\bibitem[Henriksen \& Mamon(1994)]{HMam94} \joref{Henriksen M.~J., Mamon G.~A.}{\apj}{421}{L63}{1994}

\bibitem[Hoyle et {al.}(2001)]{Hoyle01} \joref{Hoyle,~F., Outram, P.~J., Shanks,~T., et al. }{\mnras}{submitted}{}{2001} \ (arXiv:astro-ph/0102163) 

\bibitem[Iovino et {al.}(1996)Iovino, Clowes \& Shaver]{IovCS96} \joref{Iovino A., Clowes R., Shaver P.}{\aaps}{119}{265}{1996} 



\bibitem[Jaffe et {al.}(2000)]{Jaffe00} \epref{Jaffe~A.~H., et {al.}}{arXiv:astro-ph/0007333}{2000}  


\bibitem[Kirilova \& Chizhov(2000)]{KirCh00} \joref{Kirilova, D.~P., Chizhov, M.~V.}{\mnras}{314}{256}{2000} 


\bibitem[Landy \& Szalay(1993)]{LS93} \joref{Landy S.~D., Szalay A.~S.}{\apj}{412}{64}{1993}

\bibitem[Lange et {al.}(2000)]{Boom00a} \epref{Lange A.~E. et {al.} (Boomerang collaboration)}{arXiv:astro-ph/0005004}{2000}
\dobold{ 
\bibitem[Lesgourgues et {al.}(2000)]{Lesg00} \joref{Lesgourgues J.}{NuclPhysB}{582}{593}{2000} 
\ {(arXiv:hep-ph/9911447)} 
\bibitem[Lesgourgues et {al.}(1998)Lesgourgues, Polarski \& Starobinsky]{Lesg98} \joref{Lesgourgues J., Polarski D., Starobinsky A.~A.}{\mnras}{297}{769}{1998}
}




\bibitem[Luminet \& Roukema(1999)]{LR99} Luminet J.-P. \& Roukema B.~F., 1999, in Theoretical and Observational Cosmology, NATO Advanced Study Institute, Carg\`ese 1998, ed. Lachi\`eze-Rey, M., Netherlands: Kluwer,  p117 ~(arXiv:astro-ph/9901364)

\bibitem[Mamon(1993)]{Mam93} Mamon G.~A., 1993, in The N-Body problem \& Gravitational Dynamics, ed. F. Combes \& E. Athanassoula (Meudon: Obs. Paris), p188, ~(arXiv:astro-ph/9308032)

\bibitem[Brandenberger \& Martin(2000)]{MB00b} \epref{Brandenberger~R.~H., Martin~J.}{arXiv:astro-ph/0005432}{2000}  

\bibitem[Meiksin et al.(1998)Meiksin, White \& Peacock]{MeikWP98} \joref{Meiksin~A., White~M., Peacock~J.~A.}{\mnras}{304}{851}{1999} 
\ {(arXiv:astro-ph/9812214)}

\bibitem[Miller et al.(1998)Miller, Nichol \& Batuski]{MillerNB01} \joref{Miller~C.~J., Nichol~R.~C., Batuski~D.~J.}{Science}{292}{2302}{2001} 
\ {(arXiv:astro-ph/0105423)}

\bibitem[Mohr et {al.}(1999)Mohr, Mathiesen \& Evrard]{Mohr99} \joref{Mohr J.~J., Mathiesen B., Evrard A.~E.}{\apj}{517}{627}{1999}

\bibitem[Osmer(1981)]{Osmer81} \joref{Osmer P.~S.}{\apj}{247}{762}{1981} 




\bibitem[Peebles(1999)]{Peeb99b} \joref{Peebles~P.~J.~E.}{\apj}{510}{531}{1999} 


\bibitem[Perlmutter et {al.}(1999)]{SCP9812} \joref{Perlmutter S. et al.}{\apj}{517}{565}{1999} ~(arXiv:astro-ph/9812133)



\bibitem[Riess et {al.}(1998)]{HzS98} \joref{Riess A.~G. et {al.}}{\aj}{116}{1009}{1998}


\bibitem[Roukema(2000)]{Rouk00c} \joref{Roukema B.~F.}{\basi}{28}{483}{2000} \ (arXiv:astro-ph/0010185)

\bibitem[Roukema(2001)]{Rouk01arc} \joref{Roukema B.~F.}{\mnras}{325}{138}{2001} \ (arXiv:astro-ph/0102099)



\bibitem[Roukema \& Mamon(2000)]{RM00} \joref{Roukema B.~F., Mamon G.~A.}{\aap}{358}{395}{2000}\ {(arXiv:astro-ph/9911413)} 

\bibitem[Roukema \& Mamon(2001)]{RM01} \joref{Roukema B.~F., Mamon G.~A.}{\aap}{366}{1}{2001}\ {(arXiv:astro-ph/0010511)} 



\bibitem[Roukema et {al.}(1999)]{RVMB99} \joref{Roukema B.~F., Valls-Gabaud D., Mobasher B., Bajtlik S.}{\mnras}{305}{151}{1999}

\bibitem[Sandage(1999)]{Sand99} \joref{Sandage~A.}{\apj}{527}{479}{1999}

\bibitem[Scott(1991)]{Scott91} \joref{Scott D.}{\aap}{242}{1}{1991}

\bibitem[Shanks et {al.}(2000)]{Shanks00} Shanks~T., Boyle~B.~J., Croom~S.~M., Loaring~N., Miller~L., Smith~R.~J., 2000, in Clustering at High Redshift, eds Mazure~A., Le~F\`evre~O., Lebrun~V., ASP \ (arXiv:astro-ph/0003206)


\bibitem[Smith et{ al.}(1998)]{2dfqz98}  Smith R.J., Boyle B.J., Shanks T., Croom S.M., Miller L., Read M., 1998, in IAU Symposium 179: New Horizons from Multi-Wavelength Sky Surveys, eds McLean B.J., Golombek D.A., Hayes J.J.E., Payne H.E., Kluwer, p348 




\bibitem[Tadros \& Efstathiou(1996)]{TadE96} \joref{Tadros H., Efstathiou G.}{\mnras}{282}{1381}{1996}

\bibitem[Tago et {al.}(2001)]{Tago01} \joref{Tago~E., Einasto~J., Einasto~M., M\"uller~V., Andernach H.}{submitted to \aj}{}{}{2001} \ (arXiv:astro-ph/0012537)

\bibitem[Tucker et {al.}(1998)Tucker, Lin \& Shectman]{LCRS98} Tucker D.~L., Lin H., Shectman S., in Wide Field Surveys in Cosmology, ed. S.~Colombi, Y.~Mellier, B.~Raban 



\bibitem[Weinberg(1972)]{Wein72} Weinberg S., 1972, Gravitation and Cosmology, New York, U.S.A.: Wiley

\bibitem[White et {al.}(1993)]{WNEF93} \joref{White S.~D.~M., Navarro J.~S., Evrard A.~E., Frenk C.~S.}{Nature}{366}{429}{1993}


\end{thebibliography}
\end{document}